\documentclass{article}
\PassOptionsToPackage{table}{xcolor}
\PassOptionsToPackage{numbers,sort&compress}{natbib}

\usepackage[final,main]{neurips_2026}

\usepackage[utf8]{inputenc}
\usepackage[T1]{fontenc}
\usepackage{hyperref}
\usepackage{url}
\hypersetup{
  pdftitle={GeneZip: Region-Aware Compression for Long-Context DNA Modeling},
  pdfauthor={Jianan Zhao, Xixian Liu, Zhihao Zhan, Xinyu Yuan, Hongyu Guo, Jian Tang},
  pdfsubject={Region-aware sequence compression for long-context DNA modeling},
  pdfkeywords={DNA modeling, sequence compression, region-aware routing}
}
\usepackage{booktabs}
\usepackage{amsfonts}
\usepackage{nicefrac}
\usepackage{microtype}
\usepackage{xcolor}
\usepackage{graphicx}
\usepackage{subcaption}

\usepackage{amsmath}
\usepackage{amssymb}
\usepackage{mathtools}
\usepackage{bm}
\usepackage{multirow}
\newcommand{\BPT}{\mathrm{BPT}}          %

\newcommand{\indicator}{\mathbf{1}}
\newcommand{\TokenMixing}{\mathrm{TokenMixing}}
\newcommand{\Encode}{\mathrm{Encode}}
\newcommand{\Decode}{\mathrm{Decode}}

\title{GeneZip: Region-Aware Compression for Long-Context DNA Modeling}

\author{%
Jianan Zhao$^{1,2,*}$,
Xixian Liu$^{1,2,*}$,
Zhihao Zhan$^{1,2}$,
Xinyu Yuan$^{1,2}$, \\
\textbf{Hongyu Guo}$^{3,4}$,
\textbf{Jian Tang}$^{1,2,5}$ \\
$^1$Mila - Qu\'ebec AI Institute \quad
$^2$Universit\'e de Montr\'eal \quad
$^3$University of Ottawa \\
$^4$National Research Council of Canada \quad
$^5$HEC Montr\'eal \\
\href{https://andyjzhao.github.io/GeneZip/}{Project page: \nolinkurl{andyjzhao.github.io/GeneZip/}}
}

\begin{document}

\maketitle
\begingroup
\renewcommand{\thefootnote}{\fnsymbol{footnote}}
\footnotetext[1]{Equal contribution. Correspondence to Jian Tang <tangjian@mila.quebec>.}
\endgroup

\begin{abstract}
Long-context DNA modeling is computationally expensive, motivating existing approaches to reduce sequence length through uniform downsampling. However, functional information and sequence redundancy are unevenly distributed across the genome, calling for finer resolution in functionally dense regions and stronger compression of redundant sequence. We introduce GeneZip, a region-aware DNA compression framework that learns where to allocate a limited token budget. Building on H-Net-style dynamic routing, GeneZip combines a Region-Aware Ratio objective, which uses regional labels derived from GENCODE gene annotations to specify compression targets during training, with bounded routing for stable token counts. At inference, it compresses raw DNA without annotations. GeneZip provides three main benefits. First, it preserves predictive utility under high compression: GeneZip-70M achieves 137.6-fold compression with a validation perplexity increase of 0.31 relative to uncompressed Mamba2-84M, and the best average rank across four DNALongBench tasks among compression models with leakage-controlled pretraining. Second, it learns region-aware allocation with emergent repeat-aware compression. Alternative regional priors yield task-dependent performance trade-offs. Without repeat-specific supervision, GeneZip compresses transposable-element-derived and tandem repeats more strongly than matched non-repeat sequence; its tandem-repeat compression preference transfers to a nine-species evaluation panel without further training. Third, it enables efficient long-context modeling: on 450-kb inputs, GeneZip-70M achieves 36.1-fold higher inference throughput than JanusDNA-7.7M on a single H100 GPU.
\end{abstract}

\section{Introduction}
Long-context DNA models are increasingly used for sequence-to-function prediction, variant effect scoring, and 3D genome modeling, yet tasks requiring hundreds of kilobases to megabase-scale context remain bottlenecked by token mixing over very long sequences \citep{enformer_avsec2021,alphagenome_avsec2025,dnalongbench_cheng2025}. Current approaches reduce model cost with efficient long-sequence backbones or smaller capacity \citep{hyenadna_nguyen2023,caduceus_schiff2024,janusdna_duan2025}, or shorten the sequence through uniform downsampling \citep{enformer_avsec2021,alphagenome_avsec2025}. These strategies make long inputs feasible while allocating resolution according to architecture instead of genome structure. Evolution shapes uneven functional information density across DNA. On the one hand, evolutionarily constrained bases are enriched in coding exons, UTRs, and promoters relative to intergenic sequence~\citep{constraint_lindbladtoh2011}. On the other hand, transposable-element expansion and tandem repetition introduce redundancy~\citep{humangenome_ihgsc2001}. This motivates finer resolution for functionally dense regions and stronger compression of redundant sequence.

Motivated by this imbalance, we propose \textbf{GeneZip}, a region-adaptive compression module that learns non-uniform token allocation over ultra-long DNA. GeneZip provides a compression interface for long-context DNA modeling: it builds on H-Net-style dynamic chunking and routing \citep{hnet_hwang2025}, while adding biological control over where the token budget is spent. Specifically, its Region-Aware Ratio (RAR) objective trains the router toward region-wise sequence compression targets derived from static gene-structure annotations, and bounded routing enforces a global token budget for stable long-context training. The region annotations are used only during compression training, so GeneZip compresses unseen raw DNA at inference time \textit{without} region labels. We summarize GeneZip's empirical advantages along three axes:

\noindent\textbf{$\bullet$ Effective compression and prediction:} GeneZip shortens the sequence processed by the backbone by 137.6-fold, with a +0.31 perplexity increase relative to uncompressed Mamba2. It achieves the best average rank across four DNALongBench downstream tasks among compression models with leakage-controlled pretraining. These tasks cover contact map prediction, eQTL prediction, enhancer--target gene prediction, and transcription-initiation signal prediction.

\noindent\textbf{$\bullet$ Region-aware compression:} Our experiments show that GeneZip achieves high, region-aware compression with emergent repeat-aware allocation. GeneZip compresses unseen sequences without annotations at inference. Post-hoc RepeatMasker/TRF analyses reveal stronger local compression of TE-derived interspersed repeats and tandem-repeat interiors than matched non-repeat sequence, without repeat supervision.

\noindent\textbf{$\bullet$ Efficient long-context modeling:} On 450-kb inputs, 70M GeneZip achieves a 36.1$\times$ inference speedup over the uncompressed JanusDNA-7.7M model~\citep{janusdna_duan2025} (Table~\ref{tab:dnalongbench_summary}; Appendix~\ref{app:table2-inference}). By reducing the effective token-mixing length, GeneZip also supports 128K-context and 636M-parameter pretraining on a single H100 80GB GPU.

\section{Related Work}
\label{sec:related}

\textbf{Dynamic chunking and adaptive tokenization.}
Most sequence models shorten raw inputs with a fixed tokenizer or a fixed pooling rule. H-Net reframes this step as end-to-end \emph{dynamic chunking}: a learned router predicts content-dependent boundaries, and token mixing is performed on the resulting shorter sequence~\citep{hnet_hwang2025}. Its ratio loss supplies a global downsampling target without external segment labels. Recent DNA tokenization methods, including MxDNA, MergeDNA, VQDNA, and LDARNet, also move beyond fixed $k$-mers or BPE by learning sequence units, token merges, vector-quantized vocabularies, or adaptive boundaries~\citep{mxdna_qiao2024,mergedna_li2025,vqdna_li2024,ldarnet_anonymous2025}. GeneZip shares this view of learning the units presented to the backbone, with a focus on extreme, controllable sequence-length compression for long-context DNA. Concurrent dnaHNet applies H-Net's hierarchical dynamic chunking to DNA, retaining its ratio loss with stage-wise global compression targets~\citep{dnahnet_shah2026}. GeneZip introduces region-specific budget supervision to control where resolution is retained, together with bounded routing to constrain token counts.

\textbf{DNA foundation models and long-context genomic prediction.}
DNA foundation models have progressed from $k$-mer/BPE Transformer encoders such as DNABERT and DNABERT-2, through multi-species Nucleotide Transformer models, to long-context architectures such as HyenaDNA, Caduceus, JanusDNA, Evo, and Evo~2~\citep{dnabert_ji2021,dnabert2_zhou2024,nucleotide_transformer_dallatorre2025,hyenadna_nguyen2023,caduceus_schiff2024,janusdna_duan2025,evo_nguyen2024,evo2_brixi2025}. A complementary line of sequence-to-function models, including Enformer, AlphaGenome, and NTv3, uses uniform downsampling to predict regulatory tracks and variant effects over hundreds of kilobases to megabase context~\citep{enformer_avsec2021,alphagenome_avsec2025,ntv3_boshar2025}. These models motivate efficient long-context modeling, but their downsampling schedules are architecture-defined rather than tied to genomic region structure or task-specific resolution needs.

\textbf{Compression for long-range genomic modeling.}
Long-context genomic models such as NTv3, Enformer, and AlphaGenome reduce sequence length through architecture-defined downsampling schedules~\citep{ntv3_boshar2025,enformer_avsec2021,alphagenome_avsec2025}. These schedules apply the same reduction across genomic regions, although the resolution needed to preserve useful information varies with region and task: coding exons occupy only $\sim$1--2\% of the human genome and are enriched for evolutionary constraint~\citep{humangenome_ihgsc2001,constraint_lindbladtoh2011}, while regulatory signal is distributed across promoters, UTRs, introns, enhancers, and distal intergenic regions in a task-dependent manner. GeneZip instead uses region-aware ratio supervision to allocate finer resolution to selected genomic regions while retaining content-dependent boundaries. Bounded routing constrains the total token count, making this allocation controllable even at high compression. Our experiments show that GeneZip achieves high, region-aware compression with emergent repeat-aware allocation and the best average rank across four DNALongBench downstream tasks among compression models with leakage-controlled pretraining.

\section{Methodology}

\begin{figure*}[t]
  \centering
  \includegraphics[width=\textwidth]{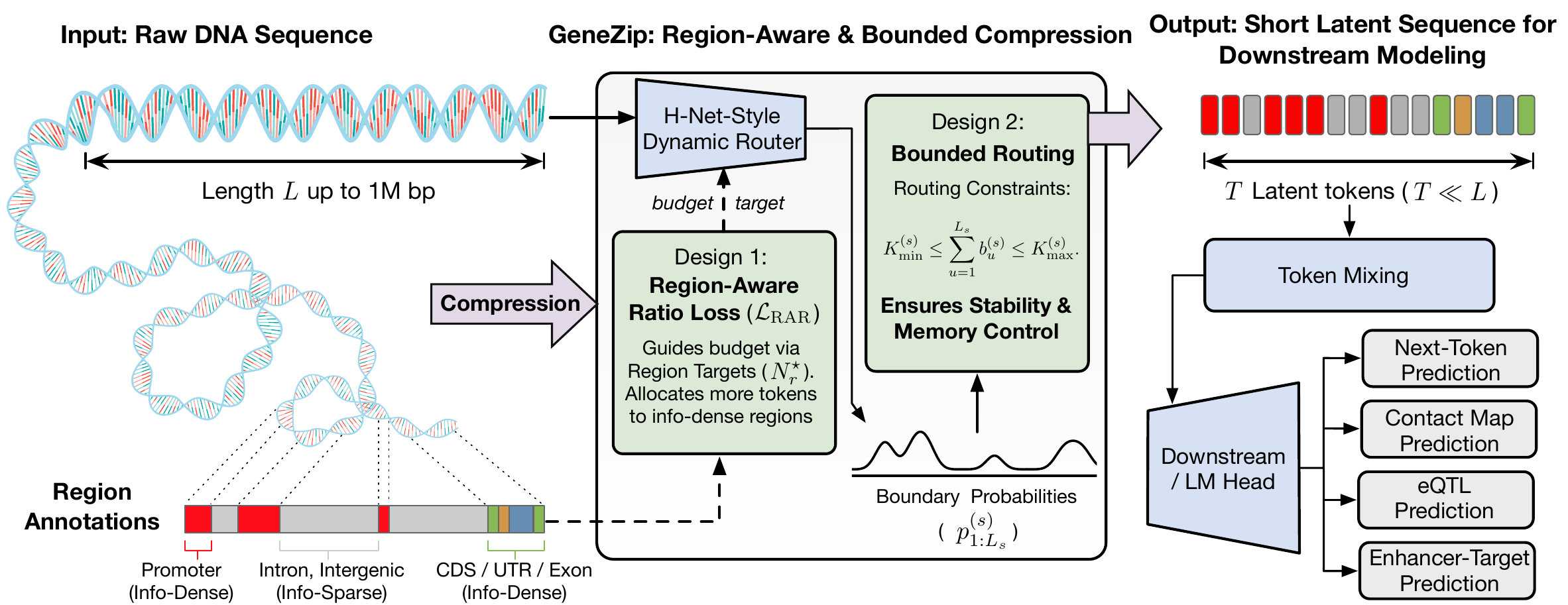}
  \caption{GeneZip overview. GeneZip compresses long-context DNA into a much shorter latent token sequence via hierarchical dynamic routing. During training, the Region-Aware Ratio (RAR) loss learns a region-adaptive token budget, and bounded routing enforces global token-count constraints for stable training. The compressed tokens are then processed by an arbitrary token-mixing backbone for downstream long-range tasks. After training, GeneZip compresses unseen raw DNA sequences \textit{without} requiring annotations or repeat masks at inference time.}
  \label{fig:genezp-overview}
\end{figure*}

\subsection{Problem setup and motivation}

Let $x_{1:L}$ denote a DNA sequence of length $L$ base pairs. Many long-range genomics tasks require
contexts on the order of $10^5$--$10^6$ bp. We decompose long-range DNA modeling into three stages---encoding, token mixing, and decoding:
\begin{equation}
\label{eq:three_stage_decomposition}
\begin{aligned}
z_{1:T} &= \Encode(x_{1:L}), \qquad T \ll L, \\
z'_{1:T} &= \TokenMixing(z_{1:T}), \\
\hat{y} &= \Decode(z'_{1:T}).
\end{aligned}
\end{equation}
Here $\Encode(\cdot)$ compresses raw DNA into a shorter latent sequence $z_{1:T}$, $\TokenMixing(\cdot)$
models interactions among the $T$ latent tokens, and $\Decode(\cdot)$ maps mixed representations to pre-training or downstream task outputs.

In long-context DNA models, the dominant compute and memory typically sit in the token-mixing backbone (e.g., Transformer layers~\citep{vaswani2017attention} or long-sequence mixers~\citep{hyenadna_nguyen2023,caduceus_schiff2024}), while encoding and decoding are comparatively lightweight. Models therefore downsample early so deep layers operate on a shorter sequence. Standard fixed-stride or pooling operators, however, allocate uniform resolution to all genomic positions~\citep{enformer_avsec2021,alphagenome_avsec2025}. This is mismatched to genomic information density: promoters, splice junctions, UTR boundaries, coding regions, regulatory motifs, and distal regulatory compartments matter differently across tasks. GeneZip addresses this by realizing $\Encode(\cdot)$ as learned non-uniform compression over long-context DNA (Figure~\ref{fig:genezp-overview}). H-Net-style dynamic routing proposes variable-length chunks~\citep{hnet_hwang2025}; the Region-Aware Ratio (RAR) loss supervises region-wise token budgets from gene-structure annotations (CDS, UTR, exon, intron, promoter, and intergenic regions). Throughout, we measure compression by base-pairs-per-token, $\BPT \coloneqq L/T$.

\subsection{GeneZip: hierarchical genomic compression with region-aware supervision}
\label{exp:GeneZip_encoder}
GeneZip instantiates $\Encode(\cdot)$ with a multi-stage hierarchical genomic compression mechanism inspired by H-Net~\citep{hnet_hwang2025}.
At each stage, the encoder (i) adaptively partitions a long sequence into variable-length chunks and (ii) selects a boundary
representation for each chunk, so that deep token mixing can be applied on a much shorter sequence. Unlike fixed-stride
pooling, the chunk boundaries are learned and depend on the input sequence.

\noindent{\textbf{Dynamic Routing module.}}
GeneZip maintains continuous sequence representations at multiple resolutions. We start from base-level embeddings
$h^{(1)}_{1:L_1}=\mathrm{Embed}(x_{1:L})$ with $L_1=L$. At routing stage $s\in\{1,\dots,S\}$, the stage-$s$ encoder $E^{(s)}$
produces per-position representations
$\hat{h}^{(s)}_{1:L_s}=E^{(s)}\!\left(h^{(s)}_{1:L_s}\right)$,
where $L_s$ is the current sequence length at stage $s$.
Following H-Net~\citep{hnet_hwang2025}, we induce adaptive boundaries by measuring the representation shift between adjacent positions in $\hat{h}^{(s)}$.
We compute boundary probabilities $p^{(s)}_t$ via cosine \emph{dissimilarity} between adjacent projected
representations:
\begin{align}
q^{(s)}_t &= W^{(s)}_q\,\hat{h}^{(s)}_t, \qquad
k^{(s)}_t = W^{(s)}_k\,\hat{h}^{(s)}_t, \\
p^{(s)}_t &=
\begin{cases}
1, & t=1,\\[2pt]
\frac{1}{2}\left(1-\dfrac{\left(q^{(s)}_{t-1}\right)^{\top} k^{(s)}_t}
{\left\|q^{(s)}_{t-1}\right\|\,\left\|k^{(s)}_t\right\|}\right), & t=2,\dots,L_s,
\end{cases} \\
\quad p^{(s)}_t & \in[0,1],  \quad \quad \tilde{b}^{(s)}_t = \indicator\left\{p^{(s)}_t > 0.5\right\}.
\end{align}
By definition $p^{(s)}_1=1$, ensuring each sequence begins with a boundary. Intuitively,
when consecutive representations span a contextual shift, their cosine similarity decreases and the
corresponding boundary probability $p^{(s)}_t$ increases.

\noindent{\textbf{Hierarchical dynamic chunking.}} At routing stage $s\in\{1,\dots,S\}$, the encoder operates on a sequence of length $L_s$ and predicts
boundary probabilities $\{p^{(s)}_t\}_{t=1}^{L_s}$ using the routing module above.
We first threshold them into a provisional boundary mask $\{\tilde{b}^{(s)}_t\}_{t=1}^{L_s}$, and then
apply bounded routing (Sec.~\ref{sec:bounded-routing}) to obtain a length-controlled final mask
$b^{(s)}=\Pi(\tilde{b}^{(s)};p^{(s)})$.
The final mask $\{b^{(s)}_t\}_{t=1}^{L_s}$ partitions the sequence into variable-length chunks. The boundary
representation of each chunk is retained, producing a shorter sequence of length
$L_{s+1}=\sum_{t=1}^{L_s} b^{(s)}_t.$
The selected sequence, denoted as $h^{(s+1)}_{1:L_{s+1}}$, serves as input to the next stage. After $S$ stages, the encoder outputs
$z_{1:T}=h^{(S+1)}_{1:L_{S+1}}$ with $T=L_{S+1}$. Compared to uniform striding, the learned
boundary policy yields non-uniform resolution: segments with frequent boundaries receive more tokens.

\noindent{\textbf{Token selection, dechunking, and gradient estimator.}}
We use the same downsampling and upsampling interface as H-Net~\citep{hnet_hwang2025}. In the downsampling path, the router's hard mask retains boundary-marked hidden states as chunk representatives; these retained states form the shorter sequence processed by the next compression stage and, after the final stage, by the token-mixing backbone. For base-level next-token pre-training, the mixed compressed sequence is decoded through the hierarchy in the reverse direction. Each compressed representation is expanded back to the positions covered by its corresponding chunk, combined with the encoder-side residual stream at that resolution, and refined by a lightweight decoder network. A nucleotide prediction head is then applied to the restored base-resolution representations. We also adopt the H-Net smoothing module, so uncertain boundaries interpolate neighboring chunk representations instead of relying on a purely hard assignment. The hard boundary mask is used in the forward pass, while the straight-through estimator and the smoothing path provide gradients to the boundary probabilities. As a result, the base-level next-token loss trains the encoder, router, token mixer, and decoder end-to-end.

\noindent{\textbf{Region supervision via gene structure.}}
To make the compression region-aware and controllable, we supervise the \emph{region-wise} token density
using gene-structure annotations. Concretely, each base position $t$ is assigned a genomic region label $r(t)\in\mathcal{R}$
(e.g. CDS, promoter, and intergenic regions). These region labels are used only to define supervision targets (RAR);
the boundary predictor itself is trained to depend only on sequence, enabling annotation-free compression at inference. Repeat labels are never used as supervision; they are used only post hoc to test whether the learned router compresses repetitive DNA more strongly than matched non-repeat sequence.

We specify a base BPT anchor $N$ and a region multiplier $\mu_r$ for each $r\in\mathcal{R}$, yielding
a region target
\begin{equation}
N_r^\star = N\cdot \mu_r.
\label{eq:region_bpt_budget}
\end{equation}
Intuitively, smaller $\mu_r$ means \emph{higher} resolution (more tokens) and larger $\mu_r$ means \emph{stronger}
compression. Promoters and genic elements are enriched for evolutionary constraint~\citep{constraint_lindbladtoh2011}, motivating finer resolution in these regions. By default, we set the multipliers to:
promoter $=1$, CDS $=1$, UTR $=2$, exon $=2$, intron $=8$, near-gene intergenic $=8$, distal-intergenic $=16$. We compare alternative pretrained allocation priors across four downstream tasks in Appendix~\ref{app:region_ratio_ablation}.

\noindent{\textbf{Region-Aware Ratio (RAR) loss.}}
RAR guides each region toward its target compression across $S$ routing stages. Because successive stages compound the reduction in sequence length, we split the region target $N_r^\star$ (Eq.~\ref{eq:region_bpt_budget}) into equal multiplicative factors. This gives each stage a target compression and corresponding fraction of positions to retain:

\begin{equation}
N_{r}^{\star(s)} \triangleq (N_r^\star)^{1/S},
\qquad
\tau_{r}^{\star(s)} \triangleq \frac{1}{N_{r}^{\star(s)}},
\qquad s=1,\dots,S.
\label{eq:stagewise_region_target}
\end{equation}

Applying these targets requires a region label for each position at every stage. The first stage uses the base-level annotations $r^{(1)}(t)\in\mathcal{R}$; subsequent stages inherit the labels of the boundary-selected representatives from the preceding stage.
Specifically, if $i_t^{(s-1)}$ indexes the $t$-th position retained by $b^{(s-1)}$, then
$r^{(s)}(t)=r^{(s-1)}(i_t^{(s-1)})$. These propagated labels are used only for RAR supervision.
For the $L_r^{(s)} \triangleq \sum_{t=1}^{L_s}\indicator\{r^{(s)}(t)=r\}$ positions assigned to region $r$, we track both the fraction actually retained, $F_r^{(s)}$, and the mean boundary probability, $G_r^{(s)}$. The first measures the realized allocation; the second provides a differentiable quantity through which the loss can adjust the router:
\begin{equation}
\label{eq:rar_region_stats}
\setlength{\jot}{2pt}
\begin{aligned}
F_r^{(s)} &\triangleq \frac{1}{L_r^{(s)}}\sum_{t=1}^{L_s} \indicator\{r^{(s)}(t)=r\}\,b_t^{(s)}, \\
G_r^{(s)} &\triangleq \frac{1}{L_r^{(s)}}\sum_{t=1}^{L_s} \indicator\{r^{(s)}(t)=r\}\,p_t^{(s)} .
\end{aligned}
\end{equation}

To adjust the boundary probabilities according to the realized allocation's deviation from the target, we couple these two statistics using the ratio loss:
\begin{equation}
\label{eq:ratio-loss-region}
\begin{split}
\mathcal{L}^{(r,s)}_{\mathrm{RAR}}
= \frac{N_r^{\star(s)}}{N_r^{\star(s)}-1}
\Big( (N_r^{\star(s)}-1)\,F_r^{(s)}\,G_r^{(s)} + (1-F_r^{(s)})(1-G_r^{(s)}) \Big).
\end{split}
\end{equation}

\noindent\emph{Intuition.}
RAR couples the realized keep rate $F_r^{(s)}$ after projection with its expectation $G_r^{(s)}$ before projection to match $\tau_r^{\star(s)}$.
When $F_r^{(s)}=G_r^{(s)}$, Eq.~\eqref{eq:ratio-loss-region} is minimized at
$F_r^{(s)}=G_r^{(s)}=\tau_r^{\star(s)}$ with $\mathcal{L}^{(r,s)}_{\mathrm{RAR}}=1$. If too many positions are retained, the loss pushes the regional boundary probabilities down; if too few are retained, it pushes them up. We weight each regional loss by its share of positions at that stage and sum across stages, so each region contributes in proportion to its representation:
\begin{equation}
\mathcal{L}_{\mathrm{RAR}}
=\sum_{s=1}^{S}\sum_{r\in\mathcal{R}} \pi_r^{(s)}\,\mathcal{L}^{(r,s)}_{\mathrm{RAR}},
\qquad
\pi_r^{(s)} \triangleq \frac{L_r^{(s)}}{L_s}.
\end{equation}

\begin{figure*}[t]
  \centering
  \setlength{\tabcolsep}{4pt}
  \renewcommand{\arraystretch}{0}

  \begin{tabular}{@{}c c c@{}}
    \includegraphics[width=0.31\textwidth,keepaspectratio]{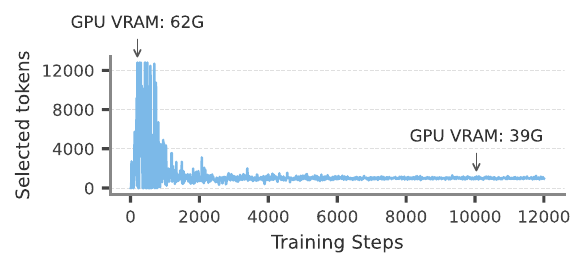} &
    \includegraphics[width=0.31\textwidth,keepaspectratio]{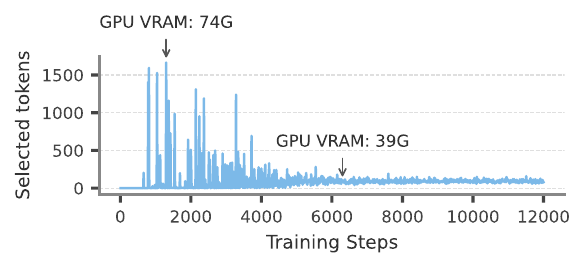} &
    \includegraphics[width=0.31\textwidth,keepaspectratio]{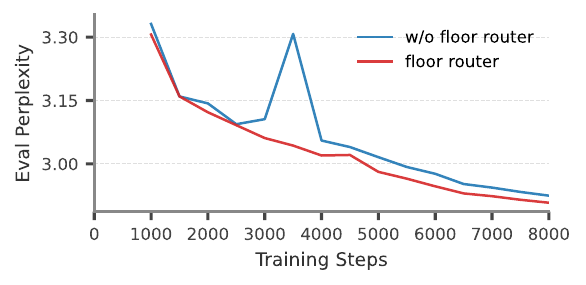} \\
    \vspace{-0.25em}
    \scriptsize (a) Selected tokens in stage 1 &
    \scriptsize (b) Selected tokens in stage 2 &
    \scriptsize (c) Eval PPL (with and without floor) \\
  \end{tabular}

  \vspace{-0.05em}
  \caption{\textbf{Why bounded routing is necessary.}
  (a) In stage 1 compression, the router can be highly unstable early on and occasionally selects an excessively large number of tokens, which risks memory blow-up; a \emph{ceiling} constraint prevents such pathological spikes.
  (b) In stage 2 compression, the router may collapse to selecting too few tokens (often near 1) at the beginning, which can induce training spikes and slow down optimization; a \emph{floor} constraint enforces a minimum token budget for stable learning.
  (c) The floor constraint yields smoother perplexity trajectories compared to training without it.}
  \label{fig:bounded_routing_diagnostics}
  \vspace{-0.75em}
\end{figure*}

\subsection{Bounded Routing for Stable Training}
\label{sec:bounded-routing}

GeneZip learns how to compress DNA through the RAR loss, progressively approaching the target regional compression ratios. During this learning process, however, the number of retained tokens can fluctuate substantially: insufficient compression can produce too many tokens and cause GPU memory spikes, whereas excessive compression can leave too few tokens for stable learning (Figure~\ref{fig:bounded_routing_diagnostics}). We therefore introduce bounded routing, which constrains the token count at each stage to a prescribed interval, using a ceiling to limit memory usage and a floor to prevent excessive token reduction.

\noindent{\textbf{Reference token budget.}}
RAR specifies how token density should vary by region; bounded routing constrains the total token count permitted at each stage. We use the base BPT anchor $N$ to define a reference keep rate $\tau^{(s)}=N^{-1/S}$, with $\prod_{s=1}^{S}\tau^{(s)}=1/N$. This reference sets the scale for the allowed interval, while the region-specific targets $N\mu_r$ continue to guide the learned allocation.
Given current length $L_s$, the reference token budget is $K^{(s)} \triangleq \tau^{(s)} L_s$.
We define a floor and ceiling around $K^{(s)}$:
\begin{equation}
K_{\min}^{(s)}=\rho_{\min}^{(s)}K^{(s)}, \qquad K_{\max}^{(s)}=\rho_{\max}^{(s)}K^{(s)}.
\end{equation}

\noindent{\textbf{Projection (bounded routing).}}
We compute a projected mask $b^{(s)}=\Pi(\tilde{b}^{(s)};p^{(s)})$ that satisfies
\begin{equation}
K_{\min}^{(s)} \le \sum_{t=1}^{L_s} b^{(s)}_t \le K_{\max}^{(s)}.
\end{equation}
The projection flips the fewest boundary decisions using $p^{(s)}$ as confidence: if too few boundaries are selected, it activates the largest-$p^{(s)}_t$ inactive positions; if too many are selected, it deactivates the smallest-$p^{(s)}_t$ active positions. The first position in each segment is always kept.

\subsection{Training and Inference}

\noindent{\textbf{Training objective.}}
Following H-Net~\citep{hnet_hwang2025}, we train GeneZip end-to-end with an autoregressive objective combined
with the multi-stage RAR regularizer.
Let $\mathcal{L}_{\mathrm{NTP}}$ denote base-level next-token cross-entropy.
The overall objective is:
$\mathcal{L}=\mathcal{L}_{\mathrm{NTP}}+\alpha\,\mathcal{L}_{\mathrm{RAR}}$, where $\alpha$ controls the strength of region-aware compression supervision.

\noindent{\textbf{Annotation-free inference.}}
At inference, GeneZip compresses unseen DNA \textit{without} annotations: sequence-predicted boundaries and bounded routing produce tokens that can be fed into any token-mixing backbone (Transformer/Mamba) and decoded for downstream tasks.

\begin{figure*}[t]
  \centering
  \vspace{-0.3em}
  \includegraphics[width=\textwidth]{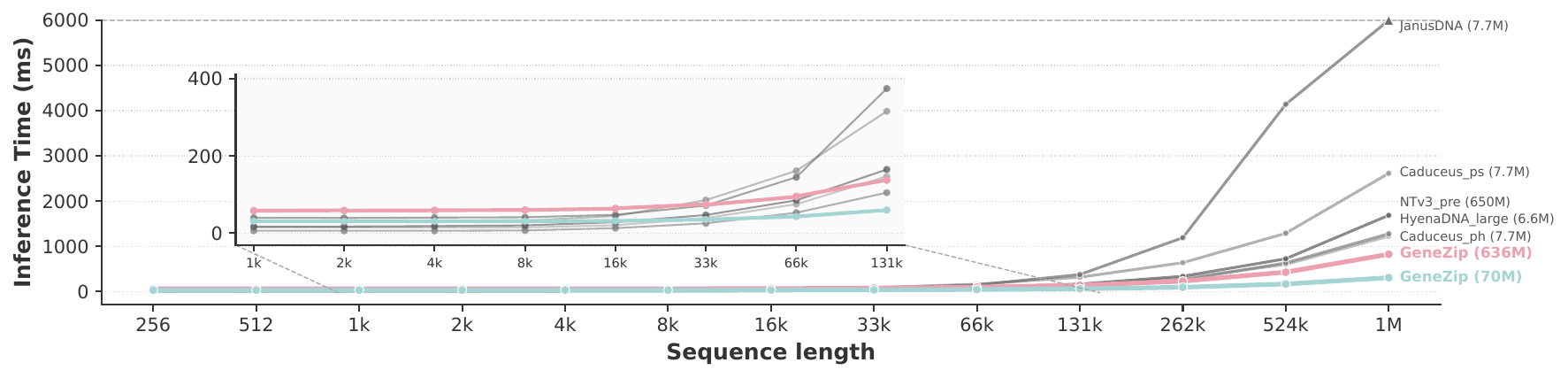}
  \vspace{-1.8em}
  \caption{\textbf{Single-sequence inference latency on long-context inputs.} Mean forward-pass latency versus input length (256 bp to 1M bp); measurement details are in Appendix~\ref{app:fig3-benchmark}. GeneZip (70M/636M) maintains low latency across the measured range. The inset shows the 1K--131K range.}
  \label{fig:inference_scaling}
  \vspace{-0.8em}
\end{figure*}

\noindent{\textbf{Inference efficiency and long-context scaling.}}
Figure~\ref{fig:inference_scaling} shows that GeneZip maintains low inference latency as input length increases to 1M bp, with a widening latency advantage over the baseline models at longer contexts. This scaling behavior supports its use for long-context DNA modeling. Benchmark details are provided in Appendix~\ref{app:fig3-benchmark}, and Section~\ref{sec:dnalongbench_downstream} reports complementary results on real genomic sequences.

\section{Experiments}
We pretrain GeneZip on 100B base pairs sampled from the human GRCh38 reference genome with a 12.8K-bp context. Regional labels are derived from GENCODE gene annotations~\citep{gencode_frankish2019} to supervise region-aware compression. We set the regional multipliers to $\mu_r=(1,1,2,2,8,8,16)$ for promoter, CDS, UTR, exon, intron, near-intergenic (NIG), and distal-intergenic (DIG) regions, respectively. Unless stated otherwise, GeneZip denotes the 70M model; the 636M and 128K suffixes identify larger-model and longer-context variants, respectively. Pretraining and data details are provided in Appendices~\ref{sec:pretrain-hparams} and~\ref{app:training_data_details}.

We evaluate whether high compression preserves useful information, how the router allocates tokens, and whether the resulting representations support long-range prediction. Section~\ref{sec:pretraining_eval} measures reconstruction quality and its response to model scaling and continued long-context pretraining; Section~\ref{sec:exp_repeat_aware} examines emergent compression of repetitive DNA; and Section~\ref{sec:dnalongbench_downstream} evaluates downstream predictive utility.

The appendix develops these findings along three complementary directions. To characterize predictive performance alongside computational cost, Appendices~\ref{app:inference-efficiency} and~\ref{app:additional_experiments} report inference scaling and speed, together with complete task-level results. Appendix~\ref{app:comp_analysis} examines GeneZip's compression behavior, including regional budget consistency and preference transfer, repeat-aware compression across species, and routing at individual loci. Appendix~\ref{app:ablation_studies} then examines how alternative regional priors change downstream trade-offs. Finally, Appendix~\ref{app:tokenization_comparison} compares GeneZip with BPE and MxDNA to assess compression and long-context prediction relative to existing tokenization methods.

\subsection{Pretraining analysis: compression quality and scaling}
\label{sec:pretraining_eval}
GeneZip and the pretraining baselines share a Mamba2 token-mixing backbone~\citep{mamba2-dao24a}, with comparable parameter counts (70M--84M) in the main comparison. Mamba2 without compression serves as the uncompressed baseline; U-Net~\citep{unet_ronneberger2015} and H-Net~\citep{hnet_hwang2025} provide uniform and learned compression baselines, respectively. Our H-Net-70M baseline follows the same dynamic-chunking and global ratio-loss formulation used by dnaHNet~\citep{dnahnet_shah2026}, instantiated with a Mamba2 backbone and a target of 128 bp/token. We train it alongside GeneZip on the same corpus, context length, and pretraining budget to compare global and region-aware compression under a common training setup. The remaining pretraining baselines also share these training conditions. Larger and longer-context GeneZip variants test scaling (Appendix~\ref{sec:pretrain-hparams}). We report overall and region-wise validation perplexity (PPL).

\begin{table*}[t]
\caption{Region-wise autoregressive perplexity (PPL) on the validation dataset. We omit \texttt{dig} (distal-intergenic) because it is weakly constrained and often heavily repeated, making PPL less informative. \textbf{Bold} indicates the best (lowest) PPL among high-compression methods (U-Net, H-Net-70M, and GeneZip). \underline{Underline} indicates the best PPL among 70M high-compression models trained at 12.8K context length.}
\label{tab:pretrain_ppl}
\vspace{-0.35em}
\centering
\small
\setlength{\tabcolsep}{4pt}
\renewcommand{\arraystretch}{1.06}
\begin{tabular}{lcccccccc}
\toprule
\textbf{Model} & \textbf{BPT}  & \textbf{Overall} $\downarrow$ & \textbf{Prom.} $\downarrow$ & \textbf{CDS} $\downarrow$ & \textbf{UTR} $\downarrow$ & \textbf{Exon} $\downarrow$ & \textbf{Intron} $\downarrow$ & \textbf{NIG} $\downarrow$ \\
\midrule
\multicolumn{9}{l}{\textit{Backbone / fine-grained upper bounds}} \\
Mamba2-84M          & 1.0 & 2.4135 & 2.0494 & 2.4136 & 2.5608 & 2.6346 & 2.5076 & 2.4826 \\
H-Net-3BPT-70M             & 3.0   & 2.4115 & 2.0694 & 2.4815 & 2.4306 & 2.8113 & 2.5462 & 2.4357 \\
\midrule
\multicolumn{9}{l}{\textit{Uniform / hierarchical compression baselines}} \\
U-Net-70M           & 128.0 & 2.8414 & 3.1335 & 3.4665 & 3.2389 & 3.0586 & 2.8415 & 2.6474 \\
H-Net-70M           & 122.2 & 2.7930 & 2.8940 & 3.4657 & 3.0241 & 3.0277 & 2.8270 & 2.6375 \\
\midrule
\multicolumn{9}{l}{\textit{GeneZip (region-aware compression)}} \\
GeneZip-70M     & 137.6 & \underline{2.7259} & \underline{2.6765} & \underline{3.4124} & \underline{\textbf{2.9084}} & \underline{2.9847} & \underline{2.7761} & \underline{2.6278} \\
GeneZip-70M-128K          & 169.2 & 2.6557 & 2.7520 & 3.0866 & 3.1236 & 3.0401 & 2.7350 & 2.5422 \\
GeneZip-636M    & 137.0 & 2.6620 & \textbf{2.6094} & \textbf{2.8272} & 2.9311 & 3.2052 & 2.7724 & 2.5741 \\
GeneZip-636M-128K    & 161.1 & \textbf{2.6484} & 2.8121 & 2.9647 & 3.1037 & \textbf{2.8712} & \textbf{2.6486} & \textbf{2.4941} \\
\bottomrule
\end{tabular}
\vspace{-0.6em}
\end{table*}

Table~\ref{tab:pretrain_ppl} establishes the reconstruction cost of high compression. Uncompressed Mamba2 and the lightly compressed H-Net-3BPT both achieve an overall PPL of approximately 2.41. At compression levels above 100 BPT, H-Net and U-Net have higher PPLs of 2.7930 and 2.8414, respectively.

GeneZip reduces this reconstruction penalty while using fewer tokens. Among the comparable-size high-compression models trained at 12.8K context, it achieves the lowest overall and all reported region-wise PPLs at the highest BPT (137.6). Its overall PPL of 2.7259 is only 0.31 above uncompressed Mamba2, improving the compression--reconstruction trade-off over H-Net and U-Net.

Overall PPL also improves with model size and continued long-context pretraining. At 12.8K context, the 636M model lowers PPL from 2.7259 to 2.6620 at a similar BPT. After a further 100B-base pretraining stage at 128K context (Appendix~\ref{sec:pretrain-hparams}), the 70M model reaches 2.6557 PPL at 169.2 BPT, narrowing the gap to Mamba2 to 0.24. GeneZip-636M-128K achieves the lowest overall PPL among the high-compression models (2.6484).

\subsection{Repeat-aware Compression}
\label{sec:exp_repeat_aware}

Repetitive DNA is a major source of sequence-level redundancy in mammalian genomes. We audit two complementary post-hoc masks: RepeatMasker TE classes (DNA transposon, LINE, LTR, and SINE), which cover interspersed TE-derived repeats, and TRF tandem repeats (TR), which capture local periodicity~\citep{bourque2018_ten_things_te,benson1999_tandem_repeats_finder}. Neither mask is used in GeneZip or H-Net training; they are used only to test whether raw-sequence compression spends fewer tokens on predictable repeat interiors. We evaluate held-out human and mouse sequences (Appendix~\ref{app:repeat_ppl_details}). For each repeat subset, we compare repeat positions with matched non-repeat positions within the same full-window context and report repeat/non-repeat ratios for both PPL and BPT. A PPL ratio below $1$ indicates easier reconstruction, whereas a BPT ratio above $1$ indicates stronger local compression.

\begin{figure*}[t]
  \centering
	\includegraphics[width=0.98\textwidth]{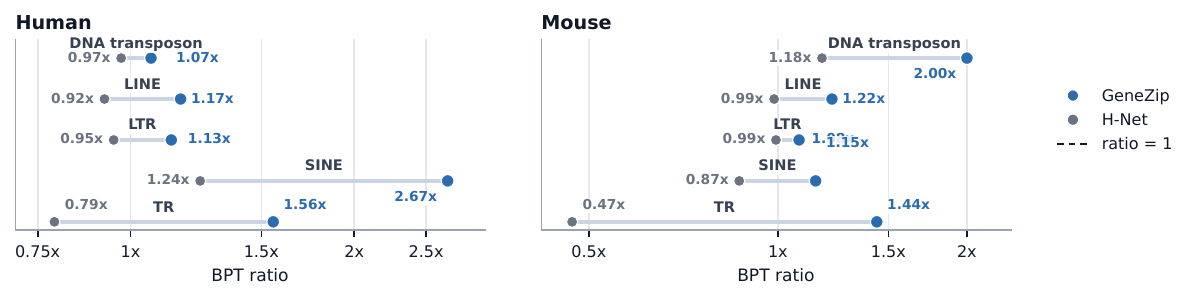}
  \vspace{-0.8em}
  \caption{\textbf{Repeat-aware compression across TE-derived and tandem-repeat sequence.}
  Repeat positions are compared with matched non-repeat positions in the same full-window context.
  BPT ratio above $1$ means stronger local compression.
  GeneZip has higher repeat/non-repeat BPT ratios than H-Net in every displayed subset, despite nearly identical repeat-subset PPL (Table~\ref{tab:repeat_ppl_by_subset}). H-Net shows weaker and less consistent repeat-specific compression.}
  \label{fig:repeat_aware_compression}
\end{figure*}
Repeat regions offer greater scope for compression because they are generally easier to predict than the surrounding background. Both GeneZip and H-Net have lower PPL on repeats than on matched non-repeat positions in nine of the ten human/mouse subsets (Table~\ref{tab:repeat_ppl_by_subset}). For tandem repeats, the repeat/non-repeat PPL ratios are 0.69--0.71 in human and 0.54--0.56 in mouse, showing that both models find these regions substantially more predictable than the background.

GeneZip exhibits an emergent \textbf{repeat-aware compression} behavior: it assigns fewer tokens per base to repetitive regions than to the background without repeat-specific supervision. Its training labels describe gene structure rather than repeat classes. Despite nearly identical repeat-subset PPL, GeneZip has higher repeat/non-repeat BPT ratios than H-Net in all ten comparisons (Figure~\ref{fig:repeat_aware_compression}). For tandem repeats, GeneZip's ratios are 1.56 in human and 1.44 in mouse, compared with 0.79 and 0.47 for H-Net: GeneZip compresses these repeats more strongly than the background, whereas H-Net compresses them less. This emergent allocation behavior also transfers beyond the human training genome, with higher TR/non-TR BPT ratios than H-Net in all nine species of a separate evaluation panel (Appendix~\ref{app:repeat_ppl_details}).

\subsection{DNALongBench downstream evaluation}
\label{sec:dnalongbench_downstream}
We evaluate whether high compression preserves downstream predictive utility by adapting the same pretrained GeneZip checkpoint separately to four DNALongBench~\citep{dnalongbench_cheng2025} tasks: contact map prediction (CMP), eQTL prediction, enhancer-target gene prediction (ETGP), and transcription initiation signal prediction (TISP). For GeneZip, Mamba2, H-Net, and U-Net, we exclude DNALongBench validation/test intervals and hold out chromosomes 8, 9, and 10 from the shared pretraining corpus. Table~\ref{tab:dnalongbench_summary} separates this leakage-controlled group from externally pretrained models and official benchmark results. For eQTL, ETGP, and TISP, GeneZip and H-Net use a common routing ceiling of 70,000 tokens per stage with no additional floor. Task- and model-specific adaptation settings and measurement protocols are provided in Appendix~\ref{app:additional_experiments}.

\begin{table*}[t]
\centering
\scriptsize
\setlength{\tabcolsep}{4.5pt}
\renewcommand{\arraystretch}{1.08}
\caption{DNALongBench results and inference cost. BPT reports validation compression; measurement protocols are detailed in Appendix~\ref{app:matched_bpt_dlb}. $^{*}$JanusDNA-7.7M uses the architectural value implied by its published single-nucleotide representation. Inference reports total time for 100 sequences of 450 kb; hardware, batching, and timing details are in Appendix~\ref{app:table2-inference}. $^{\dagger}$Evo2 runs out of memory on one GPU. Average rank averages the four descending task ranks among all 10 rows, using displayed scores and average ranks for ties; inference time is excluded. Best task scores and lowest average rank within \emph{Leakage-controlled-pretraining} are \textbf{bold}; best across all rows are \underline{underlined}.}
\label{tab:dnalongbench_summary}
\resizebox{\textwidth}{!}{%
\begin{tabular}{@{}lccccccc@{}}
\toprule
\textbf{Model} & \textbf{BPT} & \textbf{Inference (s)} $\downarrow$ & \textbf{CMP SCC} & \textbf{eQTL AUROC} & \textbf{ETGP AUPRC} & \textbf{TISP PCC} & \textbf{Avg. rank} $\downarrow$ \\
\midrule
\multicolumn{8}{@{}l}{\emph{DNALongBench-official}} \\
HyenaDNA-6.6M\citep{hyenadna_nguyen2023} & 1.0 & 22.86 & 0.115 & 0.514 & 0.249 & 0.132 & 8.00 \\
Caduceus-Ph-7.7M\citep{caduceus_schiff2024} & 1.0 & 35.06 & 0.133 & 0.566 & 0.380 & 0.109 & 5.63 \\
Caduceus-PS-7.7M\citep{caduceus_schiff2024} & 1.0 & 68.36 & 0.127 & 0.538 & 0.376 & 0.108 & 7.00 \\
\midrule
\multicolumn{8}{@{}l}{\emph{External-pretraining}} \\
JanusDNA-7.7M\citep{janusdna_duan2025} & 1.0$^{*}$ & 228.63 & \underline{0.159} & \underline{0.840} & 0.438 & 0.002 & 3.25 \\
NTv3-100M-pre\citep{ntv3_boshar2025} & 127.99 & 13.27 & 0.136 & 0.792 & \underline{0.440} & 0.194 & \underline{2.75} \\
Evo2-1B\citep{evo2_brixi2025} & 1.0 & ---$^{\dagger}$ & 0.018 & 0.754 & 0.095 & -0.002 & 9.25 \\
\midrule
\multicolumn{8}{@{}l}{\emph{Leakage-controlled-pretraining}} \\
Mamba2-84M\citep{mamba2-dao24a} & 1.0 & 40.80 & 0.111 & 0.762 & 0.322 & 0.294 & 6.25 \\
U-Net-70M\citep{unet_ronneberger2015} & 127.99 & 4.85 & 0.131 & 0.807 & \textbf{0.390} & 0.178 & 4.00 \\
H-Net-70M\citep{hnet_hwang2025} & 122.19 & 11.78 & 0.118 & 0.791 & 0.225 & 0.303 & 5.75 \\
\rowcolor{teal!10}
GeneZip-70M & 137.59 & 6.34 & \textbf{0.133} & \textbf{0.817} & 0.375 & \underline{\textbf{0.336}} & \textbf{3.13} \\
\bottomrule
\end{tabular}%
}
\end{table*}

\paragraph{Performance.}
At 137.6 BPT, GeneZip achieves the best average rank among compression models with leakage-controlled pretraining (3.13; Table~\ref{tab:dnalongbench_summary}). It leads this comparison on TISP PCC (0.336 versus 0.294 for uncompressed Mamba2) and ranks second overall to externally pretrained NTv3 (3.13 versus 2.75), with roughly twice its inference speed.

\paragraph{Efficiency.}
The separate inference benchmark measures processing cost on 100 real 450K-bp inputs, independently of task adaptation. GeneZip is $1.86\times$ faster than H-Net, while U-Net is faster still (4.85 versus 6.34 seconds in total). U-Net is close to GeneZip on CMP (0.131 versus 0.133 SCC), but has lower TISP PCC (0.178 versus 0.336).

\paragraph{Regional-prior ablation.}
Different downstream tasks have different biological focuses. The promoter-focused Default (Promoter+Genic) and PromoterDistal priors both perform well on TISP (mean PCC 0.3356 and 0.3377; Appendix~\ref{app:region_ratio_ablation}). GeneZip lets users choose regional pretraining targets to match the biological focus of their downstream task.

\section{Conclusion}
We propose GeneZip, a region-aware DNA compressor that combines bounded dynamic routing with a region-aware compression objective to reallocate token budget across genomic compartments. GeneZip achieves 137.6$\times$ compression with a 0.31 perplexity increase relative to uncompressed Mamba2, preserves downstream utility across long-context genomic tasks, and achieves the best average rank across four DNALongBench downstream tasks among compression models with leakage-controlled pretraining. Without repeat-specific supervision, GeneZip exhibits emergent repeat-aware compression, allocating fewer tokens per base to TE-derived and tandem-repeat DNA than to the surrounding non-repeat sequence. GeneZip also supports 128K-context and 636M-parameter training, while its 70M model achieves a 36.1$\times$ inference speedup over JanusDNA-7.7M for 450-kb inputs on a single H100 GPU. These results demonstrate high, region-aware compression with emergent repeat-aware allocation and support GeneZip as an efficient compression interface for long-context DNA modeling.

\clearpage
\begin{ack}
This work was supported by the Natural Sciences and Engineering Research Council (NSERC) Discovery Grant, the Canada CIFAR AI Chair Program, and an NRC Collaborative R\&D Project (AI4D-CORE-132). This work was also partially funded by the IVADO Fundamental Research Project grant PRF-2019-3583139727. Computational resources were provided by Mila, Calcul Québec, the Digital Research Alliance of Canada, and the National Research Council Canada.
\end{ack}

\bibliographystyle{unsrtnat}
\bibliography{references}

\newpage
\appendix
\section{Implementation Details}
\subsection{Pretraining Settings}
\label{sec:pretrain-hparams}
\label{sec:exp_pretraining_settings}
We pretrain on GENCODE-derived human corpora at 12.8K and 128K bp context lengths. Each window stores the raw sequence together with a run-length encoded segmentation over the 7-region taxonomy used throughout the paper (promoter, CDS, UTR, exon, intron, near-intergenic (NIG), and distal-intergenic (DIG)). To reduce DNALongBench leakage, we exclude benchmark validation/test intervals from \texttt{train}/\texttt{valid} sampling and hold out \texttt{chr8}, \texttt{chr9}, and \texttt{chr10}; full construction details and corpus statistics are provided in Appendix~\ref{app:training_data_details} and Table~\ref{tab:data_stats}.

We use a two-stage hierarchical dynamic chunking encoder with $S=2$, where each stage is implemented with two Mamba2 layers. The token-mixing backbone consists of 15 Mamba2 layers. We pretrain GeneZip with region-aware compression using this 7-region taxonomy. For the GeneZip model, we fix the base BPT anchor to $N=32$ bp/token across both stages. The GeneZip model focuses on genic information, with RAR objective targets over the seven regions (promoter:CDS:UTR:exon:intron:NIG:DIG $= 1:1:2:2:8:8:16$); Appendix~\ref{app:region_ratio_ablation} studies alternative $\mu_r$ settings for task-aware allocation. For bounded routing, we apply a shared absolute floor across all GeneZip models with $K_{\min}^{(s)}=8$ for all stages $s$, meaning at least the top-8 tokens are kept in each routing layer. The default 70M model uses unrestricted upper routing budgets; the 636M model uses ceilings of $K_{\max}^{(1)}=30{,}000$ and $K_{\max}^{(2)}=10{,}000$. Unless otherwise noted, all stages of GeneZip models and baseline models are trained on a total budget of 100B base pairs each. All reported experiments except Evo2-1B can be run on a single NVIDIA A100 80GB or H100 GPU; Evo2-1B downstream adaptation uses multi-GPU parallelism because of its model size and sharding constraints; the short-context inference-only compression measurement uses one A100.

\paragraph{Pretraining baselines.}
We train comparable-size models (70M--84M) on 100B base pairs of GENCODE data with a 12.8K context. The U-Net baseline replaces GeneZip's encoder with an autoregressive U-Net~\citep{unet_ronneberger2015}. H-Net-70M uses a high-compression target of 128 bp/token; H-Net-3BPT-70M provides a low-compression reference following~\citet{hnet_hwang2025}. Mamba2 provides an uncompressed reference~\citep{mamba2-dao24a}.

\paragraph{Phase 1 (12.8K) pretraining.}
We train on \texttt{gencode\_human\_12.8k} with a maximum sequence length of $12{,}800$ bp.
We optimize with AdamW using learning rate $1\times 10^{-3}$, weight decay $0.1$,
$(\beta_1,\beta_2)=(0.9,0.95)$, a linear learning-rate schedule with $500$ warmup steps,
and gradient clipping to $\lVert g \rVert_2 \le 2.0$.
We use per-device batch size $32$ with gradient accumulation $8$ (effective batch size $256$) and train for 100B base pairs using BF16 mixed precision.
We evaluate using $3{,}000$ validation samples.

\paragraph{Phase 2 (128K) pretraining.}
We continue training on \texttt{gencode\_human\_128k} with a maximum sequence length of $128{,}000$ bp,
initializing from the 12.8K checkpoint introduced above.
We keep the same optimizer and schedule (AdamW; LR $1\times 10^{-3}$; $500$ warmup steps; weight decay $0.1$;
$(\beta_1,\beta_2)=(0.9,0.95)$; gradient clipping $2.0$; BF16).
We use per-device batch size $4$ with gradient accumulation $8$ (effective batch size $32$) and train for another 100B base pairs using BF16 mixed precision.
We evaluate using $3{,}000$ validation samples.

\begin{table*}[t]
\caption{Pre-training hyperparameters for the GeneZip models.}
\label{tab:pretrain-hparams}
\centering
\small
\setlength{\tabcolsep}{6pt}
\begin{tabular}{lcc}
\toprule
 & 12.8K pre-train & 128K pre-train \\
\midrule
Dataset & \texttt{gencode\_human\_12.8k} & \texttt{gencode\_human\_128k} \\
Region loss weight $\alpha$ & 0.03 & 0.03 \\
Max length (bp) & $12{,}800$ & $128{,}000$ \\
Per-device batch size & $32$ & $4$ \\
Grad.\ accumulation & $8$ & $8$ \\
Update steps & $30{,}600$ & $24{,}414$ \\
Optimizer & AdamW & AdamW \\
Learning rate & $1\times 10^{-3}$ & $1\times 10^{-3}$ \\
LR schedule & Linear + warmup & Linear + warmup \\
Warmup steps & $500$ & $500$ \\
Weight decay & $0.1$ & $0.1$ \\
Adam $(\beta_1,\beta_2)$ & $(0.9,0.95)$ & $(0.9,0.95)$ \\
Grad clip & $2.0$ & $2.0$ \\
Precision & BF16 & BF16 \\
Validation samples & $3{,}000$ & $3{,}000$ \\
\bottomrule
\end{tabular}
\end{table*}

\subsection{Existing assets and licenses/terms.}
\label{app:licenses}
We cite the original sources for all third-party datasets, codebases, and public checkpoints used in the experiments. GENCODE v49 and the mouse GENCODE-derived annotations are open-access GENCODE project data.\footnote{\url{https://www.gencodegenes.org/pages/data_access.html}} The DNALongBench task data used here (CMP, eQTL, ETGP, and TISP) are distributed through Harvard Dataverse under CC0 1.0, and the DNALongBench code is released under BSD-3-Clause.\footnote{\url{https://github.com/ma-compbio/DNALONGBENCH}; task data DOIs are listed in the official README.} The public HyenaDNA, Caduceus, JanusDNA, and Evo2 code repositories are released under Apache-2.0 where applicable; their checkpoints, and the Nucleotide Transformer resources, are used under the corresponding public model-card or repository terms (Nucleotide Transformer: CC BY-NC-SA 4.0). We use these third-party assets only for research benchmarking and do not redistribute the original DNALongBench datasets or public baseline checkpoints in this work.

\section{Inference Efficiency}
\label{app:inference-efficiency}

\subsection{Length scaling}
\label{app:fig3-benchmark}

We measure how inference latency scales with input length, using a fixed batch size of one (Figure~\ref{fig:inference_scaling}).
We measure the forward-pass latency on a single NVIDIA A100 80GB GPU with BF16 inference.
For each context length $L \in \{256,512,\ldots,1{,}048{,}576\}$ bp, we construct a length-$L$ DNA string by repeating a fixed
sequence snippet and encode it into single-nucleotide token IDs.

\paragraph{Timing protocol.}
We time only the model forward pass, excluding model loading, input construction and host-to-device transfer.
For each $(\text{model}, L)$ pair, we run $2$ warm-up forward passes, followed by $10$ timed runs.
Each timed run synchronizes the device before and after the forward pass to record wall-clock time.
We report the mean of the per-forward latency in milliseconds in Figure~\ref{fig:inference_scaling}.

\subsection{Inference speed}
\label{app:table2-inference}
This section reports the total time to process 100 real genomic sequences of 450K bp using model-specific batch sizes. Table~\ref{tab:dnalongbench_summary} reports processing time, and Table~\ref{tab:inference_resources} reports the corresponding batch sizes and peak allocated GPU memory.

We use the same first 100 alternate sequences from the \textit{Artery\_Tibial} test split on one NVIDIA H100 80GB GPU with BF16, retaining the dataset's orientation, variant masking and N padding without truncation. We evaluate pretrained backbones with mean pooling and newly initialized binary classification heads, independently of task-specific adaptation. Inputs are pretokenized in pinned CPU memory.

Timing includes host-to-device transfer, mask construction, the backbone forward pass, pooling and classification; it excludes checkpoint loading, tokenization, compilation, warmup, output transfer to CPU and numerical checks. We report the median of three complete passes after warmup, synchronizing CUDA before and after each pass. Each model uses the largest batch size verified on all 100 inputs, including a final partial batch; outputs are checked against batch-1 inference with absolute and relative tolerances of 0.02. 

In this inference-only benchmark, GeneZip uses routing ceilings of $(70{,}000,70{,}000)$, and H-Net has no routing limits; these timing conditions are separate from the common downstream routing settings in Appendix~\ref{app:regional-prior-settings}. NTv3 pads inputs to 450,048 positions, masks the added positions from pooling, and includes padding cost in timing. Peak allocated GPU memory includes model weights and is measured at each model's reported batch size. 

GeneZip processes the 100 inputs in 6.34 seconds, compared with 11.78 seconds for H-Net and 228.63 seconds for JanusDNA-7.7M, corresponding to $1.86\times$ and $36.1\times$ speedups. U-Net is faster at 4.85 seconds, while GeneZip achieves a better average rank across the four downstream tasks (3.13 versus 4.00; Section~\ref{sec:dnalongbench_downstream}).

\begin{table}[tbp]
\centering
\footnotesize
\caption{Batch sizes and peak GPU memory for the inference speed benchmark reported in Table~\ref{tab:dnalongbench_summary}. Measurements use one H100 80GB GPU with BF16. Each batch size is verified on all 100 inputs, with the final partial batch included in timing. Peak allocated memory includes model weights and is measured at the reported batch size. Dashes indicate unavailable measurements under this setup.}
\label{tab:inference_resources}
\setlength{\tabcolsep}{4pt}
\begin{tabular}{@{}lrrr@{}}
\toprule
Model & Size (M) & Batch size & Peak memory (GiB) \\
\midrule
HyenaDNA & 6.6 & 6 & 41.15 \\
Caduceus-Ph & 7.7 & 19 & 75.04 \\
Caduceus-PS & 7.7 & 12 & 73.16 \\
JanusDNA-7.7M & 7.7 & 9 & 41.11 \\
NTv3-100M-pre & 100 & 7 & 56.34 \\
Evo2-1B & 1{,}000 & --- & --- \\
Mamba2 & 84 & 5 & 61.34 \\
U-Net & 70 & 47 & 70.00 \\
H-Net-70M & 70 & 5 & 62.16 \\
GeneZip & 70 & 7 & 72.80 \\
\bottomrule
\end{tabular}
\end{table}

\section{Complete Results for DNALongBench}
\label{app:additional_experiments}

\paragraph{Task coverage.}
\label{app:rsap_not_reported}
We report eQTL, ETGP, TISP, and CMP. Regulatory sequence activity prediction (RSAP) is excluded because the released data and configuration differ from the published task specification in sequence length, mouse split completeness, and training loss~\citep{dnalongbench_cheng2025}. These discrepancies prevent a directly comparable evaluation under the published protocol.\footnote{Task specification and configuration: \url{https://github.com/ma-compbio/DNALONGBENCH}; released data: \url{https://doi.org/10.7910/DVN/MNUEZR}.}

\paragraph{Shared comparison protocol.}
\label{app:dlb_shared_settings}
Stronger compression leaves fewer tokens to represent predictive information and can reduce downstream accuracy. For a fair comparison, we compare GeneZip, U-Net, and H-Net at similar validation compression around 128 BPT, using the same leakage-controlled pretraining corpus and the same DNALongBench task data, splits, and official input lengths. GeneZip achieves slightly higher compression: 137.6 BPT, compared with 128.0 for U-Net and 122.2 for H-Net (Table~\ref{tab:dnalongbench_summary}). We use the default GeneZip checkpoint from Appendix~\ref{sec:pretrain-hparams} across all four tasks; alternative regional priors are evaluated in Appendix~\ref{app:region_ratio_ablation}. Table~\ref{tab:dnalongbench_summary} distinguishes these models from official benchmark results and externally pretrained models.

For eQTL, ETGP, and TISP, the three models share AdamW learning rate $2\times10^{-4}$, weight decay $0.1$, per-device batch size 1, and gradient clipping $\lVert g\rVert_2\le1.0$. We freeze the compression frontend and train the subsequent backbone modules and task head. For GeneZip and H-Net, this freezes the embeddings, encoder, and router; for U-Net, it freezes the embeddings, stem, and downsampling blocks while leaving the bottleneck and upsampling blocks trainable. CMP instead trains a contact-map head on fully frozen sequence representations (Appendix~\ref{subsec:cmp_dlb}).

\paragraph{Compression measurement.}
\label{app:matched_bpt_dlb}
Table~\ref{tab:dnalongbench_summary} reports pretraining-validation BPT. For GeneZip ($137.5858$), H-Net ($122.1925$), and Mamba2, BPT is computed from the pretraining-validation evaluation in Table~\ref{tab:pretrain_ppl} as total valid input bases divided by total final-stage hard token counts. For U-Net, NTv3, HyenaDNA, both Caduceus variants, and Evo2-1B, we measure BPT on the first 3,000 validation windows using BF16 and batch size 1. Each 12,800-bp window contributes 12,799 input bases after the causal next-token shift; sequences are uppercased and no special tokens are added. We count bottleneck positions for U-Net, positions entering the first compressed transformer layer for NTv3, and final sequence-representation positions for uncompressed backbones. Fixed-downsampling padding is included in the token count but excluded from the base count: U-Net and NTv3 emit 100 tokens per window, giving $127.99$ BPT. JanusDNA-7.7M's $1.0$ BPT is inferred from its published single-nucleotide representation and position-wise fusion~\citep{janusdna_duan2025}; directional copies are not additional sequence positions. Please note that all these BPT values describe pretraining-validation sequences; downstream compression also depends on task inputs and routing settings.

\paragraph{Regional-prior evaluation and routing settings.}
\label{app:regional-prior-settings}
The matched-BPT regional-prior study measures the four GeneZip checkpoints and H-Net on a common 3,000-window pretraining-validation panel (Table~\ref{tab:region_ratio_ablation}). On this panel, GeneZip and H-Net have BPT values of $139.3533$ and $122.4828$, respectively. This is a separate measurement from the pretraining-validation statistics in Table~\ref{tab:pretrain_ppl}.

Table~\ref{tab:region_ratio_ablation} reports held-out test metrics. H-Net, U-Net, and all four GeneZip priors follow the common adaptation protocols and original DNALongBench sequence lengths in Appendix~\ref{app:dlb_shared_settings}; CMP uses the frozen-representation protocol in Appendix~\ref{sec:cmp-hparams}. eQTL reports AUROC and AUPRC averaged over nine tissues.

For eQTL, ETGP, and TISP, H-Net and all four GeneZip priors use the same ceiling of 70,000 tokens at each routing stage, during both adaptation and validation/test evaluation. No additional routing floor is applied; content-dependent routing and the first-valid-token rule remain enabled. The ceiling limits the number of retained tokens rather than imposing a fixed token count.

\subsection{eQTL prediction}
\label{subsec:eqtl_dlb}
We evaluate long-range variant effect modeling on the nine DNALongBench~\citep{dnalongbench_cheng2025} eQTL classification tasks. Table~\ref{tab:eqtl_auroc_models_by_dataset} reports held-out test AUROC for each tissue and the unweighted mean across nine tissues.

\label{sec:eqtl-hparams}
\label{app:classification-adaptation}
eQTL uses 450K-bp inputs, mean pooling, three training epochs, gradient accumulation 8, and BF16. AdamW uses $(\beta_1,\beta_2)=(0.9,0.95)$ with a linear learning-rate schedule and warmup ratio $0.1$. Validation and checkpoint saving occur every 100 optimizer steps; the checkpoint with highest validation AUROC is evaluated on the held-out test split. eQTL inputs preserve the benchmark's orientation, variant masking, and N padding, and test predictions use the dataset-provided orientation.

\begin{table}[tbp]
\centering
\caption{DNALongBench eQTL results across nine tissues (test AUROC; higher is better). Evaluation protocols are detailed in Appendix~\ref{sec:eqtl-hparams}. The best and second-best scores in each column are \textbf{bold} and \underline{underlined}, respectively.}
\label{tab:eqtl_auroc_models_by_dataset}

\footnotesize
\setlength{\tabcolsep}{2.6pt}
\renewcommand{\arraystretch}{1.08}

\resizebox{\columnwidth}{!}{%
\begin{tabular}{l cccccccccc}
\toprule
\multirow{2}{*}{\textbf{Model}} & \multicolumn{10}{c}{\textbf{eQTL Datasets and Average (AUROC)}} \\
\cmidrule(lr){2-11}
 & \textbf{Artery} & \textbf{Adipose} & \textbf{Fibro} & \textbf{Muscle} & \textbf{Nerve} & \textbf{Skin (NSE)} & \textbf{Skin (SE)} & \textbf{Thyroid} & \textbf{Blood} & \textbf{Average} \\
\midrule
\multicolumn{11}{l}{\emph{Task-specific baselines}} \\
Expert model\citep{dnalongbench_cheng2025} & 0.741 & 0.736 & 0.639 & 0.621 & 0.683 & 0.710 & 0.700 & 0.612 & 0.689 & 0.681 \\
CNN\citep{dnalongbench_cheng2025} & 0.576 & 0.551 & 0.547 & 0.502 & 0.516 & 0.499 & 0.499 & 0.487 & 0.577 & 0.528 \\
\midrule
\multicolumn{11}{l}{\emph{DNALongBench-official}} \\
HyenaDNA\citep{dnalongbench_cheng2025} & 0.479 & 0.513 & 0.584 & 0.487 & 0.511 & 0.471 & 0.544 & 0.529 & 0.512 & 0.514 \\
Caduceus-Ph\citep{dnalongbench_cheng2025} & 0.547 & 0.541 & 0.597 & 0.538 & 0.588 & 0.586 & 0.574 & 0.527 & 0.594 & 0.566 \\
Caduceus-PS\citep{dnalongbench_cheng2025} & 0.536 & 0.519 & 0.549 & 0.523 & 0.552 & 0.529 & 0.541 & 0.547 & 0.542 & 0.538 \\
\midrule
\multicolumn{11}{l}{\emph{External-pretraining}} \\
JanusDNA-7.7M\citep{janusdna_duan2025} & \textbf{0.852} & \textbf{0.769} & \underline{0.802} & \textbf{0.864} & \textbf{0.914} & \textbf{0.903} & \textbf{0.846} & \textbf{0.793} & \underline{0.821} & \textbf{0.840} \\
NTv3-100M-pre\citep{ntv3_boshar2025} & 0.805 & 0.744 & 0.698 & 0.798 & 0.885 & \underline{0.886} & 0.727 & \underline{0.778} & 0.808 & 0.792 \\
Evo2-1B\citep{evo2_brixi2025} & 0.745 & 0.725 & 0.691 & 0.802 & 0.816 & 0.831 & 0.705 & 0.698 & 0.771 & 0.754 \\
\midrule
\multicolumn{11}{l}{\emph{Leakage-controlled-pretraining}} \\
Mamba2 & 0.750 & 0.729 & 0.699 & 0.806 & 0.847 & 0.820 & 0.713 & 0.701 & 0.792 & 0.762 \\
U-Net & 0.845 & \underline{0.764} & 0.724 & \underline{0.837} & 0.892 & 0.876 & 0.830 & 0.715 & 0.776 & 0.807 \\
H-Net-70M\citep{hnet_hwang2025} & 0.821 & 0.721 & 0.701 & 0.792 & 0.834 & 0.884 & \underline{0.834} & 0.751 & 0.779 & 0.791 \\
\rowcolor{teal!10}
GeneZip & \underline{0.846} & 0.730 & \textbf{0.818} & 0.831 & \underline{0.902} & 0.881 & 0.785 & 0.712 & \textbf{0.849} & \underline{0.817} \\
\bottomrule
\end{tabular}%
}
\begin{minipage}{\columnwidth}
\footnotesize
\textit{Note.} Due to resource limitations, Evo2-1B was fine-tuned using LoRA on 4$\times$H100 GPUs; GeneZip runs used one A100 or H100 GPU per run, and the remaining models were fine-tuned on a single A100 GPU.
\end{minipage}
\end{table}

Among the models with leakage-controlled pretraining in Table~\ref{tab:eqtl_auroc_models_by_dataset}, GeneZip achieves the highest mean eQTL AUROC (0.8172), outperforming uncompressed Mamba2 across all nine tissues (Table~\ref{tab:eqtl_auroc_models_by_dataset}). These results show that high compression can preserve information useful for long-range variant-effect prediction.

\subsection{Enhancer-target gene prediction}
\label{subsec:etgp_dlb}

We assess regulatory interaction prediction on the DNALongBench enhancer-target gene prediction (ETGP) task~\citep{dnalongbench_cheng2025}, using CRISPRi\_EPI\_K562\_hg19. The benchmark restricts candidate enhancers to within 450 kbp of the target TSS. We use the same 450K-bp inputs, mean pooling, training schedule, and validation-AUROC checkpoint selection as for eQTL (Appendix~\ref{sec:eqtl-hparams}).

ETGP reveals a different preference for where resolution is retained. GeneZip improves AUPRC over region-agnostic H-Net (0.3746 versus 0.2251), but uniform-downsampling U-Net is slightly stronger (0.3896; Table~\ref{tab:etgp_auroc_auprc_singlecol}). Enhancer--target prediction links distal regulatory sequence with a target promoter, so prioritizing promoter and genic detail can leave useful enhancer context coarsely represented. The regional-prior comparison supports this interpretation: GeneZip-Regulome and GeneZip-PromoterDistal retain finer resolution in broader noncoding regions and achieve AUPRC of 0.5049 and 0.4893, respectively (Appendix~\ref{app:region_ratio_ablation}). These gains show why the default promoter/genic preference need not be the strongest allocation for regulatory-interaction prediction. Adjusting the prior provides a way to preserve the sequence context required by the task.

\begin{table}[t]
\caption{DNALongBench enhancer-target gene prediction results (test AUROC and AUPRC). Higher is better. Best scores are in \textbf{bold}.}
\label{tab:etgp_auroc_auprc_singlecol}
\centering
\small
\setlength{\tabcolsep}{6pt}
\renewcommand{\arraystretch}{1.08}
\begin{tabular}{lcc}
\toprule
\textbf{Model} & \textbf{AUROC} & \textbf{AUPRC} \\
\midrule
\multicolumn{3}{l}{\emph{Task-specific baselines}} \\
Expert Model\citep{dnalongbench_cheng2025}   & \textbf{0.926} & 0.407 \\
CNN\citep{dnalongbench_cheng2025}            & 0.797          & 0.310 \\
\midrule
\multicolumn{3}{l}{\emph{DNALongBench-official}} \\
HyenaDNA\citep{dnalongbench_cheng2025}       & 0.828          & 0.249 \\
Caduceus-Ph\citep{dnalongbench_cheng2025}    & 0.826          & 0.380 \\
Caduceus-PS\citep{dnalongbench_cheng2025}    & 0.821          & 0.376 \\
\midrule
\multicolumn{3}{l}{\emph{External-pretraining}} \\
JanusDNA-7.7M\citep{janusdna_duan2025}       & 0.822         & 0.438 \\
NTv3-100M-pre\citep{ntv3_boshar2025}       & 0.823         & \textbf{0.440} \\
Evo2-1B\citep{evo2_brixi2025}       & 0.737         & 0.095 \\
\midrule
\multicolumn{3}{l}{\emph{Leakage-controlled-pretraining}} \\
Mamba2\citep{mamba2-dao24a}                 & 0.783         & 0.322 \\
U-Net & 0.849 & 0.390 \\
H-Net-70M & 0.688 & 0.225 \\
\rowcolor{teal!10}
GeneZip-70M & 0.823 & 0.375 \\
\bottomrule
\end{tabular}
\vspace{-0.6em}
\end{table}

\subsection{Transcription initiation signal prediction}
\label{subsec:tisp_dlb}
We evaluate nucleotide-resolution transcription initiation signal prediction (TISP) on DNALongBench~\citep{dnalongbench_cheng2025}, predicting 10 strand-specific tracks across five assays (Table~\ref{tab:tisp_results}).

TISP uses 100K-bp inputs, a common per-base regression head, 5{,}000 updates, and gradient accumulation 4. Training uses a pseudo-Poisson KL loss with the ratio loss disabled. H-Net and all GeneZip priors use the shared 70,000-token stage ceilings without additional floors (Appendix~\ref{app:regional-prior-settings}). Checkpoints are selected by the five-assay mean PCC on the validation chromosome using the official evaluation. The larger Evo2-1B model makes full fine-tuning memory-intensive, so we use LoRA to reduce its memory requirements.

On held-out chromosomes, we score the central 50 kb of overlapping 100 kb windows and compute PCC for each assay after combining both strands, then report the mean across the five assays.

\begin{table*}[t]
\centering
\small
\setlength{\tabcolsep}{5pt}
\renewcommand{\arraystretch}{1.05}
\caption{DNALongBench transcription initiation signal prediction across five assays~\citep{dnalongbench_cheng2025}. We report Pearson correlation (PCC) for FANTOM CAGE (FC), ENCODE CAGE (EC), ENCODE RAMPAGE (ER), GRO-cap (GC), and PRO-cap (PC), along with the average. Task-specific and DNALongBench-official baselines are copied from~\citet{dnalongbench_cheng2025}; all other language-model rows follow the official TISP evaluation protocol with chromosome-level post-hoc evaluation. Best overall scores are in \textbf{bold}; best non-expert scores are \underline{underlined}.}
\resizebox{0.78\textwidth}{!}{%
\begin{tabular}{lcccccc}
\toprule
\textbf{Model} & \textbf{FC} & \textbf{EC} & \textbf{ER} & \textbf{GC} & \textbf{PC} & \textbf{Avg} \\
\midrule
\multicolumn{7}{l}{\emph{Task-specific baselines}} \\
Expert Model\citep{dnalongbench_cheng2025} & \textbf{0.808} & \textbf{0.710} & \textbf{0.749} & \textbf{0.624} & \textbf{0.774} & \textbf{0.733} \\
CNN\citep{dnalongbench_cheng2025}          & 0.029 & 0.038 & 0.043 & 0.037 & 0.066 & 0.042 \\
\midrule
\multicolumn{7}{l}{\emph{DNALongBench-official}} \\
HyenaDNA\citep{dnalongbench_cheng2025}     & 0.138 & 0.124 & 0.118 & 0.112 & 0.168 & 0.132 \\
Caduceus-Ph\citep{dnalongbench_cheng2025}  & 0.114 & 0.088 & 0.088 & 0.097 & 0.154 & 0.109 \\
Caduceus-PS\citep{dnalongbench_cheng2025}  & 0.113 & 0.088 & 0.090 & 0.102 & 0.156 & 0.108 \\
\midrule
\multicolumn{7}{l}{\emph{External-pretraining}} \\
JanusDNA-7.7M\citep{janusdna_duan2025}          & 0.003 & -0.001 & 0.003 & 0.000 & 0.003 & 0.002 \\
NTv3-100M-pre\citep{ntv3_boshar2025}       & 0.220 & 0.148 & 0.149 & 0.170 & 0.284 & 0.194 \\
Evo2-1B (LoRA)\citep{evo2_brixi2025}       & -0.002 & -0.002 & -0.004 & 0.000 & 0.000 & -0.002 \\
\midrule
\multicolumn{7}{l}{\emph{Leakage-controlled-pretraining}} \\
Mamba2\citep{mamba2-dao24a}                & 0.336 & 0.248 & 0.240 & 0.264 & 0.382 & 0.294 \\
U-Net\citep{unet_ronneberger2015} & 0.159 & 0.142 & 0.143 & 0.168 & 0.279 & 0.178 \\
H-Net-70M\citep{hnet_hwang2025} & 0.341 & 0.261 & 0.249 & 0.271 & 0.393 & 0.303 \\
\rowcolor{teal!10}
GeneZip-70M & \underline{0.392} & \underline{0.301} & \underline{0.286} & \underline{0.288} & \underline{0.411} & \underline{0.336} \\
\bottomrule
\end{tabular}%
}
\label{tab:tisp_results}
\end{table*}

GeneZip achieves a mean TISP PCC of 0.3356, exceeding H-Net (0.3032), U-Net (0.1781), and uncompressed Mamba2 (0.294; Table~\ref{tab:tisp_results}). It leads the non-expert models in Table~\ref{tab:tisp_results} across all five assays. A plausible biological explanation is that transcription initiation depends on locally organized promoter features whose positions influence where transcription begins, as demonstrated by Puffin~\citep{puffin_dudnyk2024}. GeneZip's regional prior favors finer token resolution around promoters, which may help preserve these position-sensitive features during compression. This interpretation is consistent with the lower TISP performance of GeneZip-Intergenic (a control model that reverses GeneZip's default promoter/genic preference), which shifts resolution away from promoters (Appendix~\ref{app:regional_allocation}).

\subsection{Contact map prediction}
\label{subsec:cmp_dlb}
\label{sec:cmp-hparams}
We then assess how different downsampling modules affect representation learning on a long-range genomics benchmark, the contact map prediction (CMP) task in DNALongBench~\citep{dnalongbench_cheng2025}. CMP is a two-dimensional regression task that maps a 1{,}048{,}576-bp DNA window to a binned chromatin contact map at 2{,}048-bp resolution. Following prior work~\citep{orca_zhou2022}, we use a 2D CNN as the CMP predictor backbone, but vary its input features: mean-pooled bin embeddings produced by different sequence models. Specifically, we compare the following representations: one-hot DNA embeddings (CNN-one-hot); Evo2-1B~\citep{evo2_brixi2025} with precomputed coverage-aware pooled bin embeddings from the frozen backbone; NTv3-100M-pre; JanusDNA-7.7M~\citep{janusdna_duan2025}; Mamba2 without compression; U-Net with uniform compression at 128 bp per token; H-Net-70M; and GeneZip.

For each cell-type subset (\texttt{HFF}, \texttt{H1hESC}, \texttt{GM12878}, \texttt{IMR90}, \texttt{HCT116}), we train the same lightweight CNN contact-map head on top of frozen sequence representations. The head uses projection dimension $128$, distance-embedding dimension $16$, a 2-layer 1D CNN with kernel size $5$ and dropout $0.1$, and a 3-layer 2D CNN with channels $[64,64,64]$, kernel size $3$, dropout $0.1$, and symmetrized outputs. We optimize with AdamW using learning rate $2\times 10^{-4}$, weight decay $0$, batch size $32$, BF16 mixed precision, and gradient clipping $\lVert g\rVert_2 \le 1.0$. We train for $5{,}000$ update steps.

\begin{table*}[t]
\centering
\scriptsize
\setlength{\tabcolsep}{2.6pt}
\renewcommand{\arraystretch}{1.0}
\caption{DNALongBench contact map prediction across five cell lines~\citep{dnalongbench_cheng2025}. We report stratum-adjusted correlation coefficient (SCC) and Pearson correlation (Corr) across five cell lines, along with the average. Best results are in \textbf{bold}; second best are \underline{underlined}.}
\resizebox{0.80\textwidth}{!}{%
\begin{tabular}{@{}lcccccccccccc@{}}
\toprule
\multirow{2}{*}{Model} &
\multicolumn{2}{c}{HFF} &
\multicolumn{2}{c}{H1hESC} &
\multicolumn{2}{c}{GM12878} &
\multicolumn{2}{c}{IMR90} &
\multicolumn{2}{c}{HCT116} &
\multicolumn{2}{c}{Avg} \\
\cmidrule(lr){2-3}\cmidrule(lr){4-5}\cmidrule(lr){6-7}\cmidrule(lr){8-9}\cmidrule(lr){10-11}\cmidrule(lr){12-13}
& SCC $\uparrow$ & Corr $\uparrow$
& SCC $\uparrow$ & Corr $\uparrow$
& SCC $\uparrow$ & Corr $\uparrow$
& SCC $\uparrow$ & Corr $\uparrow$
& SCC $\uparrow$ & Corr $\uparrow$
& SCC $\uparrow$ & Corr $\uparrow$ \\
\midrule
\multicolumn{13}{@{}l}{\emph{Task-specific baselines}} \\
CNN-one-hot & 0.0291 & 0.0571 & 0.0010 & 0.0297 & 0.0086 & 0.0321 & 0.0065 & 0.0302 & 0.0190 & 0.0203 & 0.0128 & 0.0339 \\
\midrule
\multicolumn{13}{@{}l}{\emph{External-pretraining}} \\
JanusDNA-7.7M & \textbf{0.1604} & \textbf{0.2125} & \textbf{0.1589} & \textbf{0.2099} & \textbf{0.1396} & \textbf{0.1872} & \textbf{0.1575} & \textbf{0.2066} & \textbf{0.1788} & \textbf{0.1993} & \textbf{0.1590} & \textbf{0.2031} \\
NTv3-100M-pre & \underline{0.1433} & \underline{0.1951} & \underline{0.1371} & \underline{0.1873} & 0.1009 & 0.1548 & \underline{0.1344} & \underline{0.1832} & \underline{0.1646} & 0.1825 & \underline{0.1361} & \underline{0.1806} \\
Evo2-1B & 0.0208 & 0.1005 & 0.0127 & 0.0918 & 0.0060 & 0.0922 & 0.0154 & 0.0912 & 0.0337 & 0.0742 & 0.0177 & 0.0900 \\
\midrule
\multicolumn{13}{@{}l}{\emph{Leakage-controlled-pretraining}} \\
Mamba2 & 0.1122 & 0.1631 & 0.1157 & 0.1657 & 0.0803 & 0.1134 & 0.1101 & 0.1588 & 0.1358 & 0.1532 & 0.1108 & 0.1508 \\
U-Net & 0.1326 & 0.1853 & 0.1306 & 0.1823 & 0.1043 & 0.1543 & 0.1269 & 0.1782 & 0.1625 & 0.1812 & 0.1314 & 0.1762 \\
H-Net-70M & 0.1130 & 0.1626 & 0.1139 & 0.1597 & 0.0917 & 0.1405 & 0.1117 & 0.1597 & 0.1606 & 0.1811 & 0.1182 & 0.1607 \\
\rowcolor{teal!10}
GeneZip & 0.1324 & 0.1840 & 0.1338 & 0.1867 & \underline{0.1071} & \underline{0.1600} & 0.1310 & 0.1804 & 0.1618 & \underline{0.1843} & 0.1332 & 0.1791 \\
\bottomrule
\end{tabular}%
}
\label{tab:cmp_results}
\end{table*}

GeneZip achieves the highest mean SCC in the leakage-controlled group (0.1332), exceeding U-Net (0.1314), H-Net (0.1182), and uncompressed Mamba2 (0.1108; Table~\ref{tab:cmp_results}). This result shows that its highly compressed representations retain information useful for long-range contact prediction. Externally pretrained JanusDNA has the highest mean SCC (0.1590), while NTv3 reaches 0.1361. GeneZip approaches NTv3's CMP performance with leakage-controlled pretraining and requires about half its inference time in the separate 450K-bp benchmark (Table~\ref{tab:dnalongbench_summary}).

\begin{samepage}
\section{Compression Analysis}
\label{app:comp_analysis}
Using checkpoints pretrained on the same leakage-controlled human corpus (Appendix~\ref{app:training_data_details}), we examine regional budget consistency and preference transfer (Section~\ref{app:regional_allocation}), repeat-aware compression without repeat supervision (Section~\ref{app:repeat_ppl_details}), and routing at representative loci (Section~\ref{sec:exp_case_study}). The case studies connect aggregate allocation patterns with local genomic features.

\end{samepage}

\begin{samepage}
\subsection{Regional Allocation and Transfer}
\label{app:regional_allocation}
\paragraph{Research questions.}
GeneZip uses region-specific targets to guide compression during training, but whether the learned compressor preserves these budgets and allocation preferences on unseen sequences remains to be established. We therefore evaluate two research questions. \textbf{RQ1 (Budget consistency)} asks whether compression remains aligned with its training-time budget targets. \textbf{RQ2 (Regional preference transfer)} asks whether the regional prior's allocation preference is preserved without annotations at inference.

\end{samepage}

We compare encoder-only checkpoints of H-Net-70M, GeneZip, GeneZip-Intergenic, and GeneZip-128K on 3{,}094 held-out human and 2{,}368 mouse windows of 128\,Kbp.\footnote{Human windows tile held-out chromosomes \texttt{chr8/chr9/chr10}; mouse windows tile a fixed-seed block of \texttt{chr11/chr19/chr10} from the GENCODE-derived mouse genome. Both use contiguous, non-overlapping windows and the standard GENCODE preprocessing ambiguity filter (\texttt{max\_n\_frac}=0.05). Of 3{,}262 human and 2{,}456 mouse windows, 168 and 88 high-\texttt{N} or assembly-gap windows are excluded, respectively.} Within each species, all checkpoints use the same windows and ambiguity filter. We obtain hard boundary masks with reverse-complement averaging and measure BPT at stage 2.

\paragraph{Metrics.}
For RQ1 (budget consistency), \emph{observed BPT} describes overall compression as input bases divided by output tokens; larger values mean coarser resolution. To assess whether this compression matches the training targets, \emph{BPT-ratio} divides observed BPT by the \emph{expected BPT} for the input's regional composition. It is equivalently the expected token count, obtained by summing regional base counts divided by their target BPTs (Eq.~\ref{eq:region_bpt_budget}), divided by the observed count; values closer to one indicate closer total-budget agreement.

Total-budget agreement can conceal offsetting regional deviations, so we also measure regional target agreement with \emph{MicroErr}. Each token is fractionally assigned to regions according to span overlap, and \emph{regional BPT} divides a region's base count by its fractional token count. MicroErr is the base-count-weighted mean absolute natural log of regional observed-to-target BPT ratios; lower values indicate smaller regional discrepancies, with zero indicating exact agreement.

For RQ2 (regional preference transfer), we ask which regions retain relatively finer resolution. BPT describes average compression within a sequence or region; to compare allocation preferences across regions, we report \emph{Token Density}, defined as each region group's tokens per base normalized by the sequence-wide average. It is the group's share of fractionally assigned tokens divided by its share of bases: values above one indicate denser allocation, and values below one indicate sparser allocation. We compare promoter, genic (CDS/UTR/exon), and intron plus intergenic (intron/NIG/DIG) groups to test whether the prior's preferred allocation direction persists.

To relate these compression measurements to reconstruction quality, we also report \emph{base-level PPL}, the exponential of mean next-base negative log-likelihood; lower values indicate better reconstruction.

\begin{table*}[t]
\caption{Compressor analysis of encoder-only checkpoints on full-chromosome held-out blocks. Human uses contiguous windows from \texttt{chr8/chr9/chr10}; mouse uses contiguous windows from \texttt{chr11/chr19/chr10}. BPT-ratio is reported as a diagnostic of global token budget; PPL and MicroErr best values are \textbf{bold} within each species. Token Density is normalized by the sequence-wide average; values above one indicate denser-than-average token allocation.}
\label{tab:pretrain_metrics}
\vspace{-0.35em}
\centering
\scriptsize
\setlength{\tabcolsep}{3pt}
\renewcommand{\arraystretch}{1.08}
\resizebox{\textwidth}{!}{%
\begin{tabular}{@{}llccccccc@{}}
\toprule
\textbf{Species}
& \textbf{Model}
& \textbf{Observed BPT}
& \textbf{BPT-ratio}
& \textbf{PPL} $\downarrow$ 
& \textbf{MicroErr} $\downarrow$ 
& \shortstack{\textbf{Intron + intergenic}\\\textbf{Token Density}}
& \shortstack{\textbf{Promoter}\\\textbf{Token Density}}
& \shortstack{\textbf{Genic}\\\textbf{Token Density}}\\
\midrule
\multirow{4}{*}{Human}
& H-Net-70M
& 124.4 $\pm$ 4.4
& 0.972 $\pm$ 0.034
& 2.8699
& -
& 1.001 $\pm$ 0.009
& 0.996 $\pm$ 0.138
& 0.970 $\pm$ 0.232 \\
& GeneZip-Intergenic
& 113.7 $\pm$ 16.1
& 1.397 $\pm$ 0.326
& 2.8618
& 0.4969
& 1.034 $\pm$ 0.038
& 0.668 $\pm$ 0.236
& 0.680 $\pm$ 0.247 \\
& GeneZip
& 189.2 $\pm$ 80.7
& 1.060 $\pm$ 0.240
& 2.8730
& 0.3372
& 0.922 $\pm$ 0.086
& 1.902 $\pm$ 1.041
& 1.593 $\pm$ 0.835 \\
& GeneZip-128K
& 198.8 $\pm$ 96.0
& 1.106 $\pm$ 0.263
& \textbf{2.8417}
& \textbf{0.2676}
& 0.891 $\pm$ 0.127
& 2.424 $\pm$ 1.872
& 1.581 $\pm$ 1.001 \\
\midrule
\multirow{4}{*}{Mouse}
& H-Net-70M
& 120.7 $\pm$ 5.2
& 0.943 $\pm$ 0.040
& 3.5559
& -
& 1.005 $\pm$ 0.011
& 0.978 $\pm$ 0.178
& 0.912 $\pm$ 0.226 \\
& GeneZip-Intergenic
& 142.2 $\pm$ 18.4
& 1.538 $\pm$ 0.420
& \textbf{3.5489}
& 0.4739
& 1.034 $\pm$ 0.041
& 0.621 $\pm$ 0.263
& 0.674 $\pm$ 0.253 \\
& GeneZip
& 162.7 $\pm$ 27.6
& 0.919 $\pm$ 0.199
& 3.5613
& 0.4189
& 0.923 $\pm$ 0.086
& 2.040 $\pm$ 1.115
& 1.682 $\pm$ 0.917 \\
& GeneZip-128K
& 186.9 $\pm$ 33.3
& 1.054 $\pm$ 0.232
& 3.5586
& \textbf{0.2566}
& 0.884 $\pm$ 0.133
& 2.832 $\pm$ 2.245
& 1.752 $\pm$ 1.270 \\
\bottomrule
\end{tabular}
}
\vspace{-0.6em}
\end{table*}

\paragraph{RQ1: Budget consistency.}
GeneZip and GeneZip-128K retain strong compression on unseen human and mouse sequences, with observed BPT of 162.7--198.8 (Table~\ref{tab:pretrain_metrics}). Their mean BPT-ratios range from 0.919 to 1.106, indicating that total compression budgets remain close to the targets after transfer. GeneZip-Intergenic shows larger total-budget deviations, with BPT-ratios of 1.397 in human and 1.538 in mouse. Total-budget adherence therefore transfers more closely for the default promoter/genic prior than for the reversed prior. At the regional level, GeneZip-128K improves target agreement relative to GeneZip: MicroErr decreases from 0.3372 to 0.2676 in human and from 0.4189 to 0.2566 in mouse.

\paragraph{RQ2: Regional preference transfer.}
Token Density reveals distinct allocation preferences despite similar reconstruction PPL within each species (Table~\ref{tab:pretrain_metrics}). H-Net-70M allocates tokens to promoters roughly in proportion to their sequence coverage, with Token Density near one. GeneZip instead gives promoters about twice its sequence-average token allocation per base (Token Density $\approx 2$), retaining finer resolution there and in genic regions; GeneZip-128K preserves this preference at longer context. GeneZip-Intergenic is a comparison model designed to reverse GeneZip's default regional preference (Appendix~\ref{app:region_ratio_ablation}). As expected, it yields promoter Token Density of 0.6--0.7 and genic Token Density below one, shifting relative resolution toward intron/intergenic sequence. Both allocation preferences persist on held-out human and mouse sequences without annotations at inference. These results show that the regional prior can be adjusted to users' preferences for where to retain finer resolution. The downstream effects of these allocation preferences are examined in Appendix~\ref{app:region_ratio_ablation}.

\subsection{Repeat-aware Compression}
\label{app:repeat_ppl_details}
We compare GeneZip and H-Net on two matched samples of approximately 10 Mb: 78 non-overlapping 128K-bp windows from held-out human chromosomes \texttt{chr8/chr9/chr10} and 78 mouse evaluation windows. RepeatMasker TE classes (DNA transposon, LINE, LTR, and SINE) and TRF masks define the evaluated repeat positions; matched non-repeat positions are drawn from the same full-window context. Tandem-repeat statistics retain windows with at least 1 kb of TRF-masked sequence and 10 kb of non-TRF sequence, leaving 70 human and 78 mouse windows. A repeat/non-repeat PPL ratio below one indicates easier reconstruction, while a BPT ratio above one indicates stronger local compression.

\paragraph{Stronger repeat compression at similar PPL.}
GeneZip compresses every evaluated repeat subset more strongly than its matched non-repeat background, and more strongly relative to that background than H-Net does (Table~\ref{tab:repeat_ppl_by_subset}). Its repeat/non-repeat BPT ratios exceed one and are higher than H-Net's in all ten human/mouse comparisons, despite nearly identical repeat-position and matched-background PPL. The strongest TE-class contrast in human is SINE, with BPT ratios of 2.67 for GeneZip and 1.24 for H-Net. For tandem repeats, GeneZip's ratios exceed one in both species, while H-Net's remain below one. Most repeat subsets also have lower PPL than the background; mouse LTR is an exception, with PPL ratios slightly above one for both models. Thus, similar reconstruction quality can coexist with substantially different token allocation: GeneZip more consistently assigns fewer tokens to repeats relative to their non-repeat background.

\paragraph{Transfer across nine species.}
To test whether this repeat-compression preference extends beyond the human training genome, we evaluate the human-trained GeneZip and H-Net checkpoints without further training on a separate RefSeq panel spanning mouse, rat, dog, cow, chicken, frog, zebrafish, Drosophila, and Arabidopsis. Each species contributes 16 sequence-deduplicated, non-overlapping 12.8K-bp windows with at most $5\%$ ambiguous bases.

Using identical windows and TRF masks, GeneZip has higher TR/non-TR BPT ratios in all nine species, with an equal-weight species mean of 1.995 versus 1.406 for H-Net (Table~\ref{tab:cross_species_repeat}). GeneZip's ratio exceeds one in all nine species, and H-Net's in eight: H-Net also compresses TR more than the background in most species of this separate panel, but GeneZip does so more strongly. Both models have TR PPL ratios below one in every species, while their PPL-ratio ordering varies. Together, the human/mouse and nine-species comparisons show that GeneZip learns stronger repeat-aware compression than H-Net without repeat-specific supervision, and that this allocation preference extends beyond the human training genome.

\begin{table*}[t]
\centering
\scriptsize
\setlength{\tabcolsep}{4pt}
\renewcommand{\arraystretch}{1.08}
\caption{\textbf{Repeat predictability and compression allocation: GeneZip versus H-Net.} Human/mouse measurements underlying Figure~\ref{fig:repeat_aware_compression}, using panels of 78 full 128K-bp windows per species; TR statistics use the 70 human and 78 mouse windows passing the support filter in Appendix~\ref{app:repeat_ppl_details}. PPL columns report mean repeat-position PPL / matched-background PPL / their ratio, averaged per window; BPT columns report mean repeat/non-repeat ratios. PPL ratio below $1$ indicates easier reconstruction and BPT ratio above $1$ indicates stronger local compression. Despite nearly identical PPL, GeneZip has higher BPT ratios than H-Net in every subset shown.}
\label{tab:repeat_ppl_by_subset}
\resizebox{\textwidth}{!}{%
\begin{tabular}{llcccc}
\toprule
\textbf{Species} & \textbf{Repeat subset} & \textbf{GeneZip repeat / bg / ratio} & \textbf{H-Net repeat / bg / ratio} & \textbf{GeneZip BPT ratio} & \textbf{H-Net BPT ratio} \\
\midrule
\multirow{5}{*}{Human} & DNA transposon & 2.67 / 2.85 / 0.94 & 2.68 / 2.85 / 0.94 & 1.07 & 0.97 \\
 & LINE & 2.17 / 2.95 / 0.74 & 2.17 / 2.95 / 0.74 & 1.17 & 0.92 \\
 & LTR & 2.28 / 2.87 / 0.80 & 2.27 / 2.87 / 0.80 & 1.13 & 0.95 \\
 & SINE & 1.69 / 2.94 / 0.58 & 1.69 / 2.94 / 0.58 & 2.67 & 1.24 \\
 & TR & 1.97 / 2.87 / 0.69 & 2.02 / 2.86 / 0.71 & 1.56 & 0.79 \\
\midrule
\multirow{5}{*}{Mouse} & DNA transposon & 2.97 / 3.58 / 0.83 & 2.98 / 3.57 / 0.83 & 2.00 & 1.18 \\
 & LINE & 3.35 / 3.60 / 0.93 & 3.35 / 3.59 / 0.94 & 1.22 & 0.99 \\
 & LTR & 3.61 / 3.58 / 1.01 & 3.61 / 3.56 / 1.02 & 1.08 & 0.99 \\
 & SINE & 3.37 / 3.59 / 0.94 & 3.37 / 3.58 / 0.94 & 1.15 & 0.87 \\
 & TR & 1.97 / 3.65 / 0.54 & 2.02 / 3.64 / 0.56 & 1.44 & 0.47 \\
\bottomrule
\end{tabular}%
}
\end{table*}

\begin{table*}[t]
\centering
\small
\setlength{\tabcolsep}{5pt}
\renewcommand{\arraystretch}{1.08}
\caption{\textbf{Tandem-repeat compression across nine species.} Human-trained models are evaluated without further training on the same 16 full 12.8K-bp windows per species. $n$ counts windows containing TR; PPL and BPT ratios compare TR with non-TR positions and are averaged over eligible windows. GeneZip has higher BPT ratios than H-Net in all nine species, while PPL-ratio ordering varies.}
\label{tab:cross_species_repeat}
\begin{tabular}{lrcccc}
\toprule
\textbf{Species} & $n$ & \textbf{GeneZip PPL ratio} & \textbf{H-Net PPL ratio} & \textbf{GeneZip BPT ratio} & \textbf{H-Net BPT ratio} \\
\midrule
Mouse & 16 & 0.539 & 0.538 & 1.571 & 0.827 \\
Rat & 16 & 0.621 & 0.621 & 2.091 & 1.783 \\
Dog & 16 & 0.625 & 0.629 & 1.753 & 1.270 \\
Cow & 16 & 0.923 & 0.782 & 1.855 & 1.488 \\
Chicken & 16 & 0.647 & 0.643 & 2.137 & 1.385 \\
Frog & 16 & 0.830 & 0.916 & 2.733 & 1.719 \\
Zebrafish & 16 & 0.635 & 0.658 & 1.986 & 1.167 \\
Drosophila & 15 & 0.928 & 0.781 & 1.645 & 1.508 \\
Arabidopsis & 11 & 0.833 & 0.852 & 2.186 & 1.510 \\
\bottomrule
\end{tabular}
\end{table*}

\subsection{Case studies of GeneZip compression}
\label{sec:exp_case_study}
\begin{figure*}[t]
  \centering
  \includegraphics[width=\textwidth]{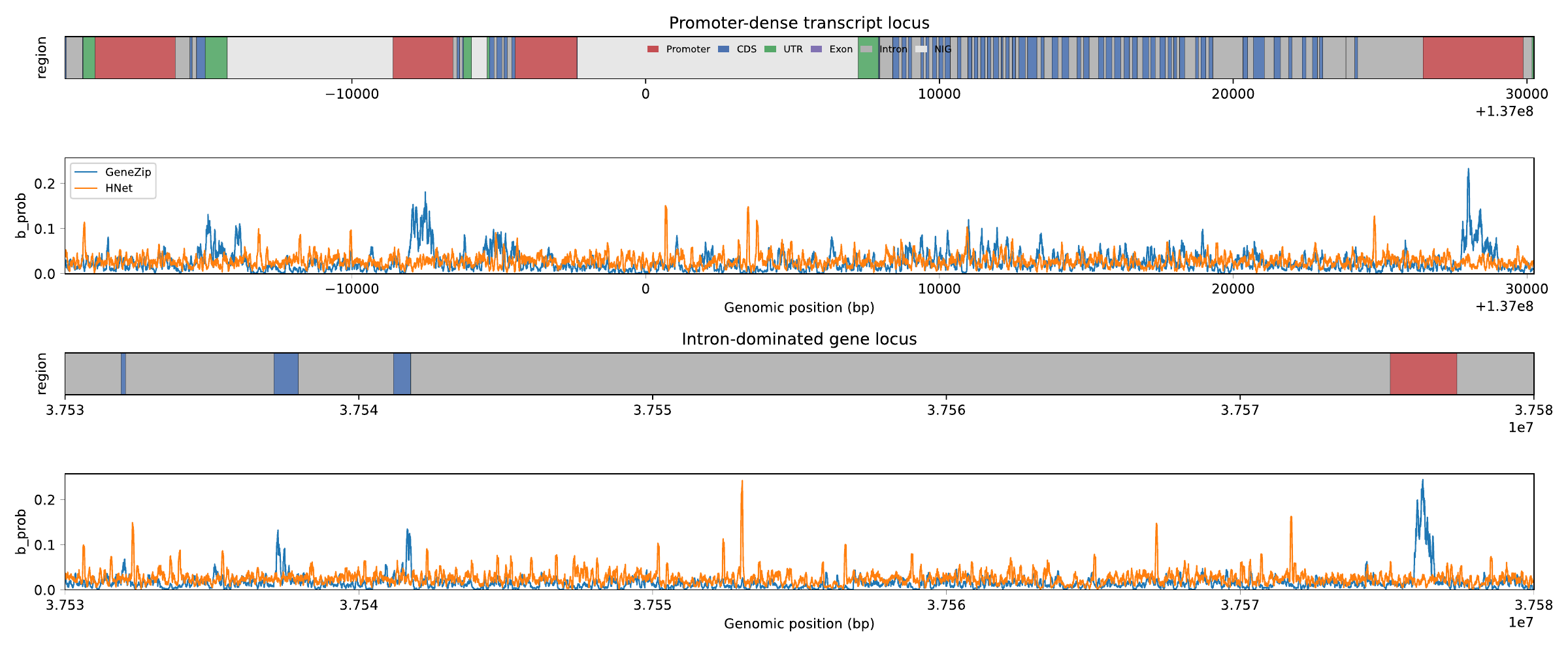}
  \vspace{-1.5em}
\caption{\textbf{Case studies of region-adaptive compression.}
Region labels are induced from GENCODE gene models \citep{gencode_frankish2019} (promoter/CDS/UTR/exon/intron/near-intergenic (NIG)), and the curves plot per-base boundary probability for H-Net \citep{hnet_hwang2025} and GeneZip.
We highlight two loci on chr9: a promoter-dense transcript locus (chr9:136{,}980{,}237--137{,}030{,}237; top) and an intron-dominated gene locus (chr9:37{,}530{,}000--37{,}580{,}000; bottom).
GeneZip shows localized boundary-probability peaks near several promoter and exonic features, with a lower background across long intronic spans.}

  \label{fig:case_study}
\end{figure*}

To make the learned routing policy more concrete, we visualize the \emph{boundary probability} (i.e., token-emission probability) along representative loci, together with a region taxonomy derived from GENCODE gene models \citep{gencode_frankish2019}. Figure~\ref{fig:case_study} overlays region labels (promoter/CDS/UTR/exon/intron/near-intergenic (NIG)) with per-base boundary probabilities produced by GeneZip ($\BPT=137.6$) and an H-Net-70M baseline ($\BPT=122.2$) \citep{hnet_hwang2025}. The plotted probabilities describe routing preferences; the final retained tokens also depend on thresholding and bounded routing.

The two loci illustrate complementary allocation patterns. In the promoter-dense locus, GeneZip raises boundary probability near several promoter and exonic features while maintaining lower values between them. In the intron-dominated locus, a long interval of low boundary probability coexists with localized peaks near annotated landmarks. H-Net has a more diffuse background with isolated spikes. These examples illustrate how the aggregate regional Token Density patterns in Appendix~\ref{app:regional_allocation} can appear locally along a sequence.

\section{Ablation Studies}
\label{app:ablation_studies}

\subsection{Regional-prior comparison}
\label{app:region_ratio_ablation}

Our default regional prior favors promoter and genic resolution because coding exons, UTRs, and promoters are enriched for evolutionarily constrained bases, reflecting the importance of preserving sequence detail in these regions~\citep{constraint_lindbladtoh2011}. We encode this preference through smaller $\mu_r$ for promoters and genic elements and larger values for intronic and intergenic sequence. Gene regulation depends on both promoters and distal regulatory elements~\citep{fantom5_forrest2014,encode_2012}, motivating comparisons with priors that redirect resolution toward broader noncoding and intergenic regions. These comparisons test how the location of retained resolution affects downstream prediction.

We compare four regional allocation priors at approximately matched pretraining-validation BPT, using one pretrained checkpoint per prior across all four downstream tasks. In the RAR objective, $N_r^\star=N\mu_r$ specifies the target BPT for region $r$; smaller $\mu_r$ preserves finer resolution. In the order (promoter, CDS, UTR, exon, intron, NIG, DIG), the multipliers are $(1,1,2,2,8,8,16)$ for GeneZip, $(1,16,2,4,2,2,4)$ for GeneZip-Regulome, $(1,16,8,8,2,2,4)$ for GeneZip-PromoterDistal, and $(32,16,8,4,2,4,1)$ for GeneZip-Intergenic. GeneZip prioritizes promoter/CDS resolution, GeneZip-Regulome emphasizes promoter and broader noncoding regulatory context, GeneZip-PromoterDistal further emphasizes promoter, intron and near-intergenic sequence, and GeneZip-Intergenic prioritizes distal intergenic sequence as a counter-prior to the default promoter/genic preference. GeneZip-Regulome and GeneZip-PromoterDistal use $N=64$; GeneZip and GeneZip-Intergenic use $N=32$. U-Net and H-Net provide uniform-downsampling and region-agnostic routing baselines, respectively.

\begin{table*}[t]
\centering
\small
\setlength{\tabcolsep}{4pt}
\renewcommand{\arraystretch}{1.08}
\caption{Regional-prior comparison across four tasks at approximately matched pretraining-validation BPT. Each GeneZip row uses one pretrained checkpoint across all tasks. GeneZip-Regulome and GeneZip-PromoterDistal use $N=64$; GeneZip and GeneZip-Intergenic use $N=32$. BPT is measured on the same 3,000 pretraining-validation windows; U-Net uses its architectural count. All scores are held-out test metrics. H-Net and all GeneZip priors use a common 70,000-token ceiling per stage without additional routing floors for eQTL, ETGP, and TISP. eQTL reports mean AUROC/AUPRC over nine tissues, ETGP reports AUROC/AUPRC, CMP reports mean SCC/Corr over five cell lines, and TISP reports mean PCC over five assays. Best scores within this table are \textbf{bold}.}
\label{tab:region_ratio_ablation}
\resizebox{\textwidth}{!}{%
\begin{tabular}{@{}lccccc@{}}
\toprule
\textbf{Model / prior} & \textbf{Validation BPT} & \textbf{eQTL AUROC/AUPRC} & \textbf{ETGP AUROC/AUPRC} & \textbf{CMP SCC/Corr} & \textbf{TISP PCC} \\
\midrule
U-Net & 127.99 & 0.8067 / 0.2019 & \textbf{0.8488} / 0.3896 & 0.1314 / 0.1762 & 0.1781 \\
H-Net-70M & 122.48 & 0.7909 / 0.1609 & 0.6881 / 0.2251 & 0.1182 / 0.1607 & 0.3032 \\
\midrule
\rowcolor{teal!10}
GeneZip & 139.35 & 0.8172 / 0.1995 & 0.8229 / 0.3746 & \textbf{0.1332 / 0.1791} & 0.3356 \\
GeneZip-Regulome ($N=64$) & 118.15 & 0.8229 / 0.1978 & 0.8290 / \textbf{0.5049} & 0.1013 / 0.1482 & 0.3278 \\
GeneZip-PromoterDistal ($N=64$) & 122.79 & \textbf{0.8241} / 0.1847 & 0.8087 / 0.4893 & 0.0932 / 0.1416 & \textbf{0.3377} \\
GeneZip-Intergenic ($N=32$) & 115.95 & 0.7865 / \textbf{0.2058} & 0.8185 / 0.4230 & 0.1280 / 0.1735 & 0.2654 \\
\bottomrule
\end{tabular}%
}
\end{table*}

Evaluation follows the shared adaptation protocols and common-panel BPT measurement described in Appendix~\ref{app:regional-prior-settings}.

\paragraph{Task-dependent biological focus.}
Different downstream tasks emphasize different biological features, motivating different regional pretraining targets. For TISP, the two promoter-focused priors, Default (Promoter+Genic) and PromoterDistal, both perform well, with mean PCC of 0.3356 and 0.3377, respectively (Table~\ref{tab:region_ratio_ablation}). The default prior leads on FC, GC, and PC, whereas PromoterDistal leads on EC and ER. Their shared emphasis on promoter resolution is consistent with the position-sensitive role of promoter features in transcription initiation~\citep{puffin_dudnyk2024}.

Tasks involving distal regulatory context favor different allocations. Regulome achieves the highest ETGP AUPRC among the GeneZip priors (0.5049), while PromoterDistal has the highest eQTL AUROC (0.8241). Both exceed the default prior on these tasks, but have lower CMP SCC (0.1013 and 0.0932 versus 0.1332 for the default). These results illustrate GeneZip's flexibility: users can choose regional pretraining targets to preserve resolution in regions relevant to the biological focus of their downstream task.

\paragraph{Biological interpretation of the reversed prior.}
GeneZip-Intergenic provides a direct contrast to the default promoter/genic preference: its learned allocation shifts resolution away from promoters and genic regions toward intron/intergenic sequence (Appendix~\ref{app:regional_allocation}). On the common pretraining-validation panel, it uses more tokens per unit sequence than GeneZip, with BPT of 115.95 versus 139.35. Yet its TISP PCC falls from 0.3356 to 0.2654, a 20.9\% reduction, with lower performance across all five assays. The greater overall token density therefore does not translate into better transcription-initiation prediction when promoter-associated resolution is reduced.

This result is consistent with the biological importance of preserving sequence detail in promoters and genic elements, where evolutionarily constrained bases are enriched~\citep{constraint_lindbladtoh2011}. Such constraint reflects purifying selection against changes that disrupt function. For transcription initiation, promoter motifs exert position-specific effects~\citep{puffin_dudnyk2024}, providing a plausible explanation for the lower TISP performance when resolution is shifted away from promoters. The GeneZip-Intergenic comparison thus suggests that retaining more tokens elsewhere cannot necessarily compensate for coarser representation of these functionally important sequences.

\section{Training Data Details}
\label{app:training_data_details}
\begin{table*}[t]
\caption{Statistics of the GENCODE-derived pretraining corpora used in this work. \textbf{Total BP} is reported in Mbp. Region columns report Mbp with within-split percentages in parentheses under our 7-region taxonomy: promoter, coding sequence (CDS), UTR, exon, intron, near-intergenic (NIG), and distal-intergenic (DIG). \textbf{N BP} reports ambiguous bases (\texttt{N}) in Mbp. The \texttt{test} split corresponds to contiguous, non-sampled windows from held-out chromosomes (\texttt{chr8}, \texttt{chr9}, \texttt{chr10}), while \texttt{train}/\texttt{valid} are sampled windows.}
\label{tab:data_stats}
\vspace{-0.35em}
\centering
\scriptsize
\setlength{\tabcolsep}{3pt}
\renewcommand{\arraystretch}{1.05}
\resizebox{\textwidth}{!}{%
\begin{tabular}{llcccccccccc}
\toprule
\textbf{Dataset} & \textbf{Split} & \textbf{Total BP (M)} & \textbf{N BP (M)} & \textbf{Promoter} & \textbf{CDS} & \textbf{UTR} & \textbf{Exon} & \textbf{Intron} & \textbf{NIG} & \textbf{DIG} \\
\midrule
\multirow{2}{*}{\textbf{GENCODE-Human-12.8k}} & train & 12160.00 & 0.16  & 1815.54 (14.93\%) & 324.84 (2.67\%) & 296.16 (2.44\%) & 190.66 (1.57\%) & 6758.93 (55.58\%) & 2636.61 (21.68\%) & 137.26 (1.13\%) \\
& valid & 640.00 & 0.01 & 96.56 (15.09\%) & 17.08 (2.67\%) & 15.51 (2.42\%) & 9.89 (1.55\%) & 355.94 (55.61\%) & 138.24 (21.60\%) & 6.79 (1.06\%) \\
\midrule
\multirow{2}{*}{\textbf{GENCODE-Human-128k}} & train & 12672.00 & 0.63 & 1047.20 (8.26\%) & 223.82 (1.77\%) & 259.47 (2.05\%) & 174.46 (1.38\%) & 7546.70 (59.55\%) & 3266.07 (25.77\%) & 154.28 (1.22\%) \\
& valid & 128.00 & 0.01 & 10.52 (8.22\%) & 2.23 (1.74\%) & 2.72 (2.12\%) & 1.79 (1.40\%) & 77.64 (60.65\%) & 32.21 (25.17\%) & 0.90 (0.70\%) \\
\midrule
\multicolumn{2}{l}{\textbf{Held-out (chr8, 9, chr10)}} & 417.34 & 17.52 & 22.44 (5.38\%) & 3.42 (0.82\%) & 4.81 (1.15\%) & 5.11 (1.22\%) & 243.77 (58.41\%) & 117.50 (28.15\%) & 20.31 (4.87\%) \\
\bottomrule
\end{tabular}%
}
\vspace{-0.6em}
\end{table*}

We construct the human pretraining corpora from the GRCh38 primary assembly together with the GENCODE v49 \textit{basic} gene annotation \citep{gencode_frankish2019}.

\paragraph{Region taxonomy and priority overlay.}
We build a genome-wide region map and label each base with exactly one region under a deterministic priority order:
\texttt{promoter} $>$ \texttt{cds} $>$ \texttt{utr} $>$ \texttt{exon} $>$ \texttt{intron} $>$ \texttt{nig} $>$ \texttt{dig}.
Promoters are defined as symmetric windows around the transcript TSS (TSS $\pm 1$ kbp).
We further partition intergenic sequence into near-intergenic (NIG) and distal-intergenic (DIG) based on proximity to annotated transcription start sites (TSSs). After applying the region-priority overlay (so that promoter/genic regions take precedence), an intergenic base is labeled as NIG if it lies within 450 kbp (genomic distance) of any annotated TSS; otherwise it is labeled as DIG.

\paragraph{Sampling, splits, and formats.}
We create two corpora with fixed context lengths: 12.8K bp and 128K bp.
For \texttt{train}/\texttt{valid}, windows are sampled with a mixture of center choices that increases coverage near transcription start sites and coding regions, alongside uniform sampling across the remaining genome.
Each window stores the raw DNA sequence together with a run-length encoded segmentation over the 7-region taxonomy.
The \texttt{valid} split is an in-distribution holdout constructed with the same sampling procedure as \texttt{train}.

\paragraph{DNALongBench leakage filtering and held-out chromosomes.}
To tailor the corpus for DNALongBench evaluation and mitigate benchmark leakage \citep{dnalongbench_cheng2025,rafi2025_homologyleakage}, we remove all genomic intervals used by DNALongBench validation/test sets from the sampling pool for \texttt{train}/\texttt{valid}.
In addition, we fully hold out the DNALongBench transcription-initiation held-out chromosomes \texttt{chr8}, \texttt{chr9}, and \texttt{chr10} during sampling.
These chromosomes are reserved to construct a chromosome-level \texttt{test} split by tiling \texttt{chr8}, \texttt{chr9}, and \texttt{chr10}  into contiguous, non-overlapping windows of the target context length (without sampling).

\section{Comparison to Tokenization Methods}
\label{app:tokenization_comparison}

Tokenization determines the units on which a DNA model operates. Viewed together with the embedding step, it serves the same role as GeneZip's encoder in Eq.~\ref{eq:three_stage_decomposition}: both map a raw nucleotide sequence to input representations for the token-mixing backbone. In this broad sense, GeneZip can be understood as end-to-end learned latent tokenization.

However, GeneZip focuses on extreme sequence-length compression for long-context modeling, whereas existing DNA tokenization methods primarily emphasize useful sequence representations. From a compression perspective, the two representative methods in Table~\ref{tab:tokenization_compression_comparison} achieve modest reductions in sequence length. BPE~\citep{bpe_sennrich2015}, whose merge rules are fixed during model training, yields 1.4 bp/token on the eQTL Artery Tibial test set. MxDNA~\citep{mxdna_qiao2024}, which learns tokenization jointly with the model, yields 3.0 bp/token on the same sequences. Even at this resolution, adapting the released MxDNA Transformer to 450K-bp eQTL inputs exceeds the available GPU memory. GeneZip supports this long-context evaluation and achieves a mean eQTL AUROC of 0.8172, compared with 0.7544 for BPE.

\begin{table}[t]
\centering
\small
\caption{Comparison of GeneZip with tokenization baselines on long-context downstream tasks. BPE and MxDNA BPT is evaluated on the eQTL Artery Tibial test set and computed as total bp divided by total tokens; the dash indicates that GeneZip has no reported BPT measurement on this panel. GeneZip task scores are from Tables~\ref{tab:eqtl_auroc_models_by_dataset} and~\ref{tab:etgp_auroc_auprc_singlecol}. BPE uses the same backbone and GENCODE training data as GeneZip; MxDNA uses the authors\textquotesingle{} released Transformer checkpoint and thus differs in backbone architecture. GeneZip follows the adaptation protocols in Appendix~\ref{app:additional_experiments}. OOM denotes an out-of-memory failure on 450K-bp inputs.}
\label{tab:tokenization_compression_comparison}
\begin{tabular}{lccc}
\toprule
\textbf{Model} & \textbf{BPT} & \textbf{eQTL AUROC} & \textbf{ETGP AUPRC} \\
\midrule
BPE & 1.3947 & 0.7544 & 0.2540 \\
MxDNA & 3.0283 & OOM & OOM \\
\rowcolor{teal!10}
GeneZip & --- & 0.8172 & 0.3746 \\
\bottomrule
\end{tabular}
\end{table}

\section{Limitations and Future Work}
\label{app:limitations_future_work}

GeneZip is currently pretrained on human genomic data. The full-chromosome human/mouse compressor analysis in Table~\ref{tab:pretrain_metrics} provides inference-time evidence that the learned routing policy transfers across held-out chromosomes and to another mammalian genome while still concentrating token budget according to the selected biological prior. Extending pretraining to broader multi-species resources with harmonized region annotations, together with downstream benchmarking across species, is an important direction for future work.

GeneZip builds on the underlying H-Net routing architecture~\citep{hnet_hwang2025}. Our contribution is to introduce a region-aware compression objective and bounded budget control on top of this dynamic routing framework. The boundary predictor still relies on representation dissimilarity between neighboring tokens, and its decisions may therefore be affected by noisy or low-complexity DNA sequences where local dissimilarity does not always correspond to functionally useful resolution.

Finally, the current supervision is centered on static region annotations and the Region-Aware Ratio objective. A promising next step is to incorporate richer function-oriented supervision beyond annotation labels, for example sequence-to-function objectives inspired by models such as Enformer and NT-family genomic predictors~\citep{enformer_avsec2021,ntv3_boshar2025}. We do not anticipate direct negative societal impact from the present methodological contribution.

\newpage
\section*{NeurIPS Paper Checklist}

\begin{enumerate}

\item {\bf Claims}
    \item[] Question: Do the main claims made in the abstract and introduction accurately reflect the paper's contributions and scope?
    \item[] Answer: \answerYes{}
    \item[] Justification: Sections~\ref{sec:pretraining_eval}, \ref{sec:exp_repeat_aware}, and \ref{sec:dnalongbench_downstream} support the compression, repeat-allocation, efficiency, and downstream claims. The downstream comparison uses a fixed regional prior and distinguishes shared-pretraining baselines from externally pretrained models; adaptation differences are described in Appendix~\ref{app:additional_experiments}.

\item {\bf Limitations}
    \item[] Question: Does the paper discuss the limitations of the work performed by the authors?
    \item[] Answer: \answerYes{}
    \item[] Justification: Appendix~\ref{app:limitations_future_work} discusses human-genome pretraining, dependence on static annotations, and reliance on H-Net routing. Appendices~\ref{app:additional_experiments} and \ref{app:region_ratio_ablation} describe the scope of the comparisons and differences in downstream adaptation.

\item {\bf Theory assumptions and proofs}
    \item[] Question: For each theoretical result, does the paper provide the full set of assumptions and a complete (and correct) proof?
    \item[] Answer: \answerNA{}
    \item[] Justification: The paper does not present theoretical guarantees or theorem-based results. Section~\ref{exp:GeneZip_encoder} defines the compression objective and routing mechanism.

\item {\bf Experimental result reproducibility}
    \item[] Question: Does the paper fully disclose all the information needed to reproduce the main experimental results of the paper to the extent that it affects the main claims and/or conclusions of the paper (regardless of whether the code and data are provided or not)?
    \item[] Answer: \answerYes{}
    \item[] Justification: The architecture and objectives are specified in Section~\ref{exp:GeneZip_encoder}; pretraining, data construction, and downstream protocols are described in Appendices~\ref{sec:pretrain-hparams}, \ref{app:training_data_details}, and \ref{app:additional_experiments}. The public code at \url{https://github.com/AndyJZhao/GeneZip} provides GeneZip implementations and reproduction instructions for pretraining and the four DNALongBench tasks.

\item {\bf Open access to data and code}
    \item[] Question: Does the paper provide open access to the data and code, with sufficient instructions to faithfully reproduce the main experimental results, as described in supplemental material?
    \item[] Answer: \answerYes{}
    \item[] Justification: GeneZip code and reproduction instructions for the main experiments---pretraining and the four DNALongBench tasks (CMP, eQTL, ETGP, and TISP)---are publicly available at \url{https://github.com/AndyJZhao/GeneZip}. The repository documents the environment, pretrained checkpoints, and commands for training and evaluation. The source datasets and data construction procedures are described in Appendices~\ref{app:licenses} and \ref{app:training_data_details}.

\item {\bf Experimental setting/details}
    \item[] Question: Does the paper specify all the training and test details (e.g., data splits, hyperparameters, how they were chosen, type of optimizer) necessary to understand the results?
    \item[] Answer: \answerYes{}
    \item[] Justification: Appendices~\ref{sec:pretrain-hparams} and \ref{app:training_data_details} specify pretraining hyperparameters, corpus construction, and leakage exclusions. Appendix~\ref{app:additional_experiments} gives task-specific training, evaluation, and checkpoint-selection settings, and Appendix~\ref{app:region_ratio_ablation} states the settings and comparability limits of the regional-prior ablation.

\newpage
\item {\bf Experiment statistical significance}
    \item[] Question: Does the paper report error bars suitably and correctly defined or other appropriate information about the statistical significance of the experiments?
    \item[] Answer: \answerNo{}
    \item[] Justification: Table~\ref{tab:pretrain_metrics} reports variation across held-out sequence windows, not across training seeds. The downstream results are point estimates without multi-seed confidence intervals or significance tests; score differences and average ranks do not establish statistical significance.

\item {\bf Experiments compute resources}
    \item[] Question: For each experiment, does the paper provide sufficient information on the computer resources (type of compute workers, memory, time of execution) needed to reproduce the experiments?
    \item[] Answer: \answerYes{}
    \item[] Justification: Appendix~\ref{sec:pretrain-hparams} describes accelerator types and pretraining budgets. Appendix~\ref{app:table2-inference} reports inference timing, batch sizes, and peak GPU memory, and Appendix~\ref{app:additional_experiments} specifies downstream hardware and training schedules.

\item {\bf Code of ethics}
    \item[] Question: Does the research conducted in the paper conform, in every respect, with the NeurIPS Code of Ethics \url{https://neurips.cc/public/EthicsGuidelines}?
    \item[] Answer: \answerYes{}
    \item[] Justification: The work uses public genomic reference resources and existing benchmark datasets for methodological research, with sources and terms described in Appendix~\ref{app:licenses}. It does not recruit participants or collect new human-subject data.

\item {\bf Broader impacts}
    \item[] Question: Does the paper discuss both potential positive societal impacts and negative societal impacts of the work performed?
    \item[] Answer: \answerNo{}
    \item[] Justification: The paper discusses the benefit of reducing the computational cost of long-context genomic modeling. Appendix~\ref{app:limitations_future_work} states that no direct negative societal impact is anticipated, but does not provide a detailed assessment of potential negative impacts or misuse.

\item {\bf Safeguards}
    \item[] Question: Does the paper describe safeguards that have been put in place for responsible release of data or models that have a high risk for misuse (e.g., pre-trained language models, image generators, or scraped datasets)?
    \item[] Answer: \answerNA{}
    \item[] Justification: The contribution is a sequence-compression method evaluated on public genomic references and prediction benchmarks, rather than a system for designing or experimentally validating biological sequences. The release does not introduce a new participant-level genomic dataset; no application-specific release safeguards are described.

\item {\bf Licenses for existing assets}
    \item[] Question: Are the creators or original owners of assets (e.g., code, data, models), used in the paper, properly credited and are the license and terms of use explicitly mentioned and properly respected?
    \item[] Answer: \answerYes{}
    \item[] Justification: The manuscript cites the datasets, annotations, codebases, and baseline models used in the experiments. Appendix~\ref{app:licenses} describes their licenses or source terms, and the public code repository includes third-party notices.

\newpage
\item {\bf New assets}
    \item[] Question: Are new assets introduced in the paper well documented and is the documentation provided alongside the assets?
    \item[] Answer: \answerYes{}
    \item[] Justification: The GeneZip code release at \url{https://github.com/AndyJZhao/GeneZip} includes an Apache-2.0 license, third-party notices, dependency specifications, and instructions for pretraining and downstream evaluation. Its README documents checkpoint variants, regional multipliers, and reproduction commands for the four DNALongBench tasks.

\item {\bf Crowdsourcing and research with human subjects}
    \item[] Question: For crowdsourcing experiments and research with human subjects, does the paper include the full text of instructions given to participants and screenshots, if applicable, as well as details about compensation (if any)?
    \item[] Answer: \answerNA{}
    \item[] Justification: The study uses existing public genomic reference resources and benchmark datasets. It does not involve crowdsourcing, participant recruitment, or new human-subject experiments.

\item {\bf Institutional review board (IRB) approvals or equivalent for research with human subjects}
    \item[] Question: Does the paper describe potential risks incurred by study participants, whether such risks were disclosed to the subjects, and whether Institutional Review Board (IRB) approvals (or an equivalent approval/review based on the requirements of your country or institution) were obtained?
    \item[] Answer: \answerNA{}
    \item[] Justification: The study does not recruit participants or collect new human-subject data. The experiments use existing public reference resources and benchmark datasets under their source terms.

\item {\bf Declaration of LLM usage}
    \item[] Question: Does the paper describe the usage of LLMs if it is an important, original, or non-standard component of the core methods in this research? Note that if the LLM is used only for writing, editing, or formatting purposes and does \emph{not} impact the core methodology, scientific rigor, or originality of the research, declaration is not required.
    \item[] Answer: \answerNA{}
    \item[] Justification: LLMs are not an important, original, or non-standard component of the core research methodology. The compression method and its training objectives are specified in Section~\ref{exp:GeneZip_encoder}.

\end{enumerate}

\end{document}